\documentclass[12pt,english]{article}
\usepackage{chapterbib}
\usepackage{titling}

\usepackage[T1]{fontenc}
\usepackage[latin9]{inputenc}
\usepackage[english]{babel}
\usepackage{amstext}

\usepackage{amsmath}
\usepackage{amsfonts}
\usepackage{amssymb}
\usepackage[normalem]{ulem}
\allowdisplaybreaks

\makeatletter
\let\start@align@nopar\start@align
\let\start@gather@nopar\start@gather
\let\start@multline@nopar\start@multline
\long\def\start@align{\par\start@align@nopar}
\long\def\start@gather{\par\start@gather@nopar}
\long\def\start@multline{\par\start@multline@nopar}
\makeatother

\usepackage[hyperindex,colorlinks,hyperfootnotes = false,citecolor = blue,]{hyperref}
\usepackage[usenames]{color}

\usepackage{graphicx}

\newcommand{\dd}{\mathrm{d}}
\newcommand\norm[1]{\left\lVert#1\right\rVert}

\newcommand{\varA}[1]{{\operatorname{#1}}}

\usepackage[top=1in, bottom=1in, left=1in, right=1in]{geometry}

\usepackage{chemarr}
\usepackage{chngcntr}

\usepackage{booktabs}
\usepackage{multirow}

\usepackage{chemarr}
\usepackage{chngcntr}

\usepackage[round,numbers,sort&compress]{natbib}

\makeatletter
\newcommand{\lyxaddress}[1]{
\par {\raggedright #1
\vspace{1.4em}
\noindent\par}
}

\makeatother

\usepackage{babel}

\usepackage{footmisc}
\DefineFNsymbols{mySymbols}{{\ensuremath\dagger}{\ensuremath\ddagger}\S\P
   *{**}{\ensuremath{\dagger\dagger}}{\ensuremath{\ddagger\ddagger}}}
\setfnsymbol{mySymbols}

\begin{document}


\title{GlnK facilitates the dynamic regulation of bacterial nitrogen assimilation}

\author{Adam Gosztolai$^{\text{}1}$
, J{\"o}rg Schumacher$^{\text{}2}$
, Volker Behrends$^{\text{}3,4}$
, Jacob G Bundy$^{\text{}4}$
,\\ Franziska Heydenreich$^{\text{}2,}\thanks{Current address: Laboratory of Biomolecular Research, Paul Scherrer Institute, 5232, Villigen, Switzerland and Department of Biology, ETH Zurich, 8093, Zurich, Switzerland}$
, Mark H Bennett$^{\text{}2}$
, Martin Buck$^{\text{}2}$
, Mauricio Barahona$^{\text{}1,}\thanks{Correspondence: m.barahona@imperial.ac.uk.}$
}

\date{}

\maketitle

\lyxaddress{1. Department of Mathematics, Imperial College London, SW7 2AZ, London, United Kingdom.\\
2. Department of Life Sciences, Imperial College London,  SW7 2AZ, London, United Kingdom.\\
3. Department of Life Sciences, University of Roehampton, SW15 5PU, London, United Kingdom.\\
4. Department of Surgery and Cancer, Imperial College London, SW7 2AZ, London, United Kingdom}

\section*{Abstract}
\paragraph*{Ammonium assimilation in \textit{E. coli} is regulated by two paralogous proteins (GlnB and GlnK), which orchestrate interactions with regulators of gene expression, transport proteins and metabolic pathways.  Yet how they conjointly modulate the activity of glutamine synthetase (GS), the key enzyme for nitrogen assimilation, is poorly understood. We combine experiments and theory to study the dynamic roles of GlnB and GlnK during nitrogen starvation and upshift. We measure time-resolved \textit{in vivo} concentrations of metabolites, total and post-translationally modified proteins, and develop a concise biochemical model of GlnB and GlnK that incorporates competition for active and allosteric sites, as well as functional sequestration of GlnK. The model predicts the responses of GS, GlnB and GlnK under time-varying external ammonium level in the wild type and two genetic knock-outs. Our results show that GlnK is tightly regulated under nitrogen-rich conditions, yet it is expressed during ammonium run-out and starvation. This suggests a role for GlnK as a buffer of  nitrogen shock after starvation, and provides a further functional link between nitrogen and carbon metabolisms.}

\section*{Introduction}

\begin{figure}[t!]
	\centering
     \includegraphics[width= 0.7\textwidth]{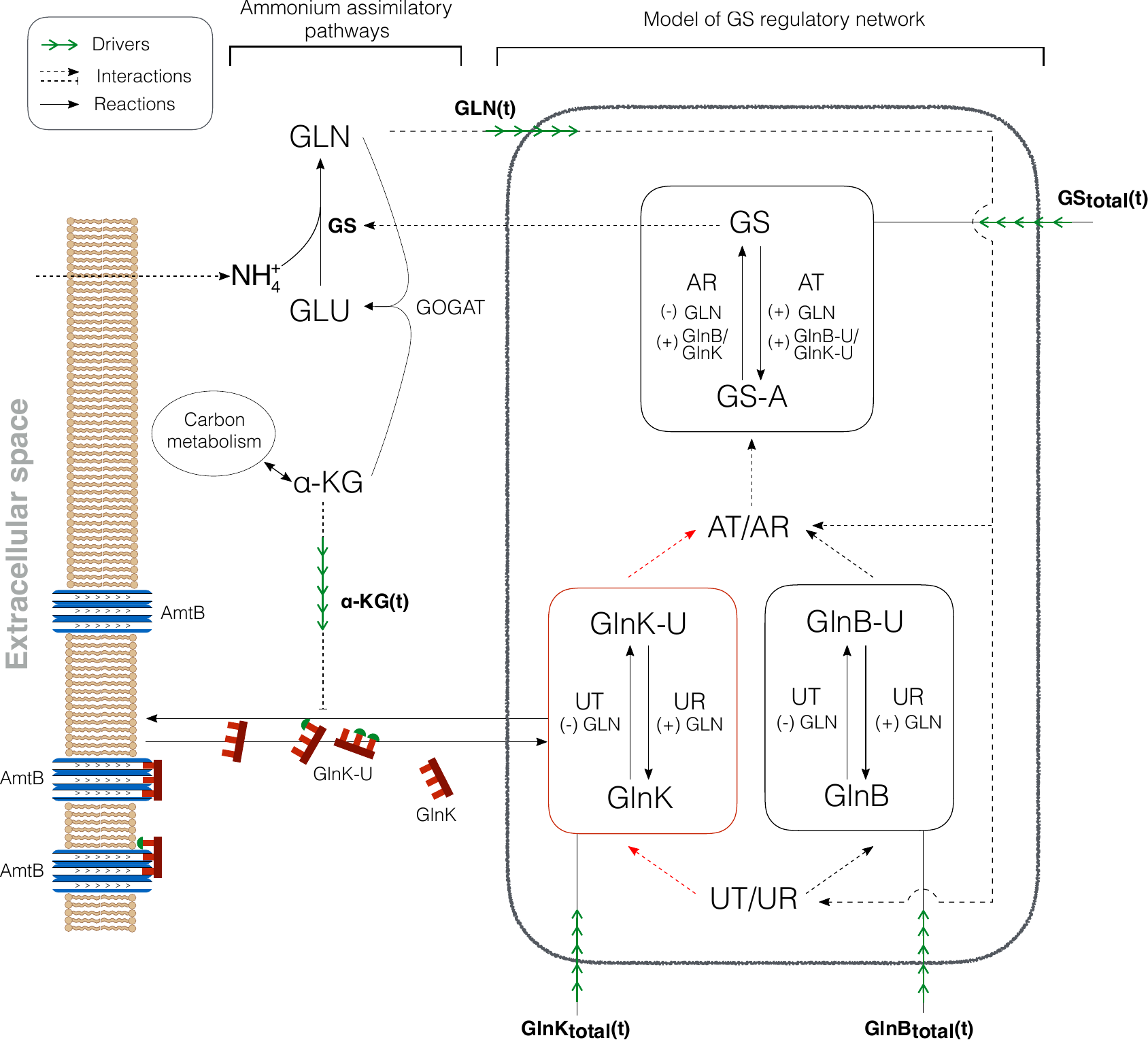}
 \caption{Summary of the ammonium assimilatory system in \textit{E. coli}. Under nitrogen limited conditions, intracellular ammonium is assimilated almost exclusively through an enzymatic reaction catalysed by GS. The activity of GS is controlled by the enzymes in the black box, which are the object of our model, as a function of the carbon ($\alpha$-KG) and nitrogen (GLN) states of the cell. Our model describes the post-translational states of GlnB, GlnK and GS, which are modified by the enzymes AT/AR and UT/UR. The total concentrations of GlnB, GlnK and GS, which reflect the transcriptional and translational responses of the cell, are measured experimentally and taken as drivers for the purposes of the model. Glutamine is sensed by UT/UR and transmitted via two complementary branches of signal transduction regulated by GlnB and GlnK. The focus of this study is GlnK (red box). Trimers of GlnK form a complex with the membrane-bound transport protein AmtB, with the effect of blocking active ammonium import. During run-out and starvation, high levels of $\alpha$-KG inhibit AmtB-GlnK complex formation, thus enabling active transport. Importantly, AmtB-GlnK complex formation sequesters GlnK, changing the amount accessible by UT/UR and AT/AR. The nitrogen and carbon information contained in the total and PTM levels of GlnB and GlnK are integrated with the level of GLN by the enzyme AT/AR to regulate the activity of GS.}
     \label{fig1}
\end{figure}

To adapt to the highly variable nutrient conditions in their natural habitat, microorganisms have evolved a complex intracellular circuitry coupling signal transduction, membrane transport, gene expression and metabolism. A widely conserved example is the ammonium assimilatory system in \textit{Escherichia coli}, which coordinates the uptake of nitrogen with other pathways involved in carbon assimilation and maintenance of cellular energy status \citep{reitzer,ninfabook,heeswijk}. External ammonium (\textit{E. coli}'s preferred nitrogen source) is assimilated into glutamate and glutamine in two steps: first, ammonium and intracellular glutamate (GLU) are converted into glutamine (GLN) in a reaction catalysed by the enzyme glutamine synthetase (GS); second, glutamine is combined with $\alpha$-ketogluterate ($\alpha$-KG), a product of carbon metabolism, to yield two molecules of glutamate (for a net production of one GLU molecule) in a reaction catalysed by the enzyme glutamate synthase (GOGAT)~(Fig.~\ref{fig1}).
    
Although glutamine and glutamate are both key intermediates towards the production of amino acids and nucleotides, their wider roles in cellular metabolism are different. Whereas glutamate is the nitrogen source for most ($88\%$) reactions in the cell \citep{reitzer}, glutamine constitutes the primary signal controlling the activity of enzymes in the ammonium assimilation system. As a result, their regulatory requirements are different: homeostatic regulation of glutamate is essential to maintain growth rate, whereas glutamine levels must adjust rapidly to reflect cellular nitrogen status while at the same time remaining within physiologically acceptable bounds. In particular, when nutrient-rich conditions are suddenly restored after a period of starvation, the cell must be able to avoid glutamine shock~\citep{,kustu,wingreen}. 

The modulation of GS activity is the primary means by which \textit{E. coli} achieves these regulatory targets in the face of changing nitrogen levels. On one hand, as extracellular ammonium is used up during growth, the activity and abundance of GS is raised so as to maintain the growth rate as long as possible. On the other hand, when favourable conditions are suddenly restored, GS can be `turned off'. This nonlinear response of GS is controlled by the bifunctional enzyme adenylyl-transferase/adenylyl-removase (AT/AR), which can reversibly adenylylate (inactivate) or deadenylylate (activate) independently each of the 12 monomers that conform GS, thus modulating GS activity in a graded manner~\citep{stadtman}. The activity of AT/AR is in turn regulated allosterically by glutamine and by two highly similar signal transduction proteins, GlnB and GlnK (encoded, respectively, by the \textit{glnB} and \textit{glnK} genes) \citep{ninfa9,ninfabook}. Importantly, GlnB and GlnK share a source of upstream regulation, themselves being reversibly uridylylated by a second bifunctional enzyme, uridilyl-transferase/uridylyl-removase (UT/UR), which enables GlnB and GlnK to control the enzyme AT/AR \citep{ninfa1}. Specifically, under glutamine limitation GlnB and GlnK are mostly uridylylated, leading to increased AR activity and inhibition of AT activity (Fig. \ref{fig1}). In addition to regulating AT/AR, GlnB and GlnK also regulate the expression of GS and bind $\alpha$-KG, communicating with carbon metabolism and signal transduction pathways \citep{ninfabook}.

Due to the apparent redundancy of having two signalling proteins (GlnB and GlnK) relaying the glutamine state of the cell, previous models~\citep{Bruggeman,doucette,ninfa3,straube} have ignored GlnK regulation of AT/AR, it being less potent than GlnB. This simplification is supported by mutational studies of the receptor interaction domains of GlnB and GlnK \citep{ninfa10} and by \textit{in vitro} studies~\citep{ninfa7}. From this viewpoint, the role of GlnK is mostly circumscribed to regulating the membrane-bound channel protein AmtB, whose gene (\textit{amtB}) lies in the same operon as \textit{glnK}, causing \textit{glnK} and \textit{amtB} expression to be directly correlated \citep{merrick3,merrick1,hwa}.

However, various experiments suggest that GlnK also acts as an effective regulator of AT/AR. For instance, GlnB deficient strains (i.e., $\Delta$\textit{glnB} mutants) exhibit unimpaired growth~\citep{ninfa8} with regulated GS (de)adenylylation \citep{Heeswijk2}, suggesting that GlnK can substitute for GlnB. It has also been found that GlnK does interact with AT/AR \textit{in vitro} \citep{ninfa7}, albeit relatively weakly, and binds $\alpha$-KG, ATP and ADP similarly to GlnB \citep{xu,ninfa13}. Importantly, \textit{glnK} expression is induced during nitrogen limitation, whereas \textit{glnB} is expressed constitutively~\citep{Heeswijk2}. Taken together, these observations raise questions about the \textit{in vivo} regulation of AT/AR and about a complementary role for GlnK in nitrogen assimilation to provide not only added redundancy but also a flexible and rapid response to the consequences of variations in external ammonium levels. 

To establish the relevance of GlnK in nitrogen regulation during external nitrogen variations, we combined experiment with modelling. We measured the growth rate (Fig.~\ref{growth}) and the \textit{in vivo} temporal response of relevant metabolites, proteins and post-translational modification (PTM)  protein states during ammonium run-out, starvation and a subsequent ammonium shock (Fig.~\ref{drivers}). Measurements of the wild type (WT) strain were complemented with experiments with two genetic knock-outs: $\Delta$\textit{glnB} (no GlnB) and $\Delta$\textit{glnK} (no GlnK), as shown in Fig.~\ref{fig_rev}. 

Our experiments show that growth rate and ammonium uptake are not impaired by the removal of either GlnB or GlnK (Fig.~\ref{growth}) suggesting that GlnB and GlnK conform two complementary branches of nitrogen signal transduction regulating AT/AR. We also find that GlnK plays a limited role in regulating GS expression and adenylylation during ammonium run-out and starvation. However,  GlnK levels rise steeply during run-out, exceeding those of GlnB four-fold during starvation (Fig.~\ref{drivers}). This suggests a functional role of GlnK, post-starvation, until normal ammonium conditions are restored. 

To confirm this hypothesis, we developed a concise, mechanistic model of GS regulation (Fig. \ref{fig1}). We show that competition between GlnB and GlnK for UT/UR active sites and AT/AR allosteric sites, as well as functional sequestration of GlnK are necessary to explain the \textit{in vivo} uridylylation and adenylylation dynamics in the WT and to predict those in the two mutants. Our model predicts the dynamic response of GS adenylylation levels in the mutants, indicating that GlnK can substitute for GlnB. Moreover, we find that GlnK is a potent regulator of AT/AR post-starvation, when GlnK levels are high relative to GlnB. Our results agree with the role of GlnK as a regulator of the ammonium transporter AmtB upon nitrogen upshift~\cite{merrick1,merrick3}, and with the decreased viability of \textit{glnK} mutants following extended ammonium starvation~\cite{blauwkamp}. Hence our work suggests a role of GlnK as an anticipatory buffer molecule to help avoid glutamine shock: GlnK is expressed before and during starvation in order to temporarily alleviate excess ammonium import and assimilation when ammonium becomes available again.

\section*{Materials and methods}
    
\subsection*{Strains and growth conditions}

Experiments were performed with \textit{E. coli} K12 (NCM3722) \cite{soupene}. NCM3722$\Delta$\textit{glnB} in-frame deletions were obtained through P1 phage transduction \cite{miller}, using strain JW2537 with the \textit{glnK} knock-out from the Keio collection as the donor strain \cite{baba}. NCM3722$\Delta$\textit{glnB} was verified by genomic locus sequencing. Pre-cultures were grown in Gutnick minimal media, supplemented with 10 mM NH\textsubscript{4}Cl, 0.4\% (w/v) glucose and Ho-Le trace elements. Main cultures were inoculated in same media but with limiting 3 mM NH\textsubscript{4}Cl, resulting in NH\textsubscript{4}Cl starvation at mid log phase. Ammonium depletion and OD600 were determined at defined time points during the course of experiments as described by \cite{Schumacher}.  Following one generation time (40 min) of growth arrest, NH\textsubscript{4}Cl was added to obtain a concentration of 3 mM (upshift).

\subsection*{Metabolite and protein measurements}

Glutamine and $\alpha$-ketoglutarate were quantified using liquid chromatography-mass spectrometry (LC-MS) and nuclear magnetic resonance (NMR), as described previously \cite{Schumacher,komorowski}. Total abundance and uridylylation level of GlnK were determined using multiple reaction monitoring mass-spectrometry (MRM-MS)~\cite{picotti} with absolute standard protein quantification (PSAQ) \cite{picard,Schumacher}. Briefly, isotopically labelled GlnK protein standard was purified following IPTG induced over-expression in \textit{E. coli} harbouring JW0440 from the ASKA(-) collection (\textit{glnK})~\cite{kitagawa}, grown in the presence of labelled L-Arginine (\textsuperscript{13}C\textsubscript{6}, \textsuperscript{15}N\textsubscript{4}) and L-Lysine (\textsuperscript{13}C\textsubscript{6}, \textsuperscript{15}N\textsubscript{2}). Ni-affinity purified GlnK standard purity was judged visually by SYPRO Ruby Protein Gel Stain (\mbox{Invitrogen}) stained SDS-PAGE to be $>$90\% and GlnK standard concentration determined. Isotopic labelling efficiency was determined by MRM-MS to be 100\%. Total GlnK abundances were determined by the ratio of MRM-MS signals of GlnK unlabelled/labelled signature peptides, excluding the uridylylation site peptide (Table S1 in the Supporting Material) and intracellular concentrations calculated~\cite{Schumacher}. GlnK uridylylation levels were directly measured by the abundance of the uridylylated GlnK peptide GAEYS(\textbf{UMP})VNFLPK compared with the non-uridylylated peptide GAEYSVNFLPK. The GlnK protein concentrations derived from non-uridylylated peptides correspond well with GlnK protein concentrations derived independently from the sum of uridylylated and non-uridylylated peptide GAEYSVNFLPK (Fig. S8 in the Supporting Material). A similar procedure is used for quantifying GlnB, GS and their PTMs~\cite{komorowski}.

\subsection*{Description of the model} 

    The schematic of the system regulating nitrogen assimilation in \textit{E. coli} is shown in Fig.~\ref{fig1}. We use three coupled non-linear ODEs to describe the dynamics of the monomeric concentrations of uridylylated GlnB, uridylylated GlnK and adenylylated GS:
    {\small
    \begin{align} \label{fluxes}
        \frac{\dd[\text{GlnB-U}]}{\dd t}&=v_\text{UT,GlnB}-v_\text{UR,GlnB} \nonumber \\
        \frac{\dd[\text{GlnK-U}]}{\dd t}&=v_\text{UT,GlnK}-v_\text{UR,GlnK}  \\
        \frac{\dd[\text{GS-A}]}{\dd t}&=v_\text{AT,GS}-v_\text{AR,GS-A} \nonumber
    \end{align}
    }
    where the right hand side contains the balance of fluxes corresponding to (de)uridylylation and (de)adenylylation of the corresponding proteins. These fluxes contain several kinetic parameters and depend non-linearly on the concentrations of substrates, products and allosteric effectors. We now describe these terms in detail.
    
\paragraph*{\textbf{Equations for UT/UR:}}
    The UT and UR activities in the bifunctional enzyme UT/UR are treated as independent, unidirectional reactions obeying Michaelis-Menten kinetics. Due to the specificity of UT/UR active sites \cite{Zhang}, GlnB and GlnK are modelled as competing substrates. Our model considers only the experimentally measured monomeric concentration of the uridylylated species, and disregard the multimeric nature of GlnB and GlnK, since (de)uridylylation of GlnB monomers is non-cooperative \cite{atkinson}.  Glutamine (GLN) acts as an allosteric activator to UR and an inhibitor to UT activities~\cite{ninfa1}. These can be encapsulated in the following rate equations (see Section S1 in the Supporting Material for derivations):
    {\small
    \begin{subequations}\label{UTUR}
    \begin{align}
        &v_\text{UT,GlnB}=\frac{V_\text{UT}\frac{[\text{GlnB}]}{K_\text{m,GlnB}}}{\left(1+\frac{[\text{GlnB}]}{K_\text{m,GlnB}}+\frac{[\text{GlnK}]}{K_\text{m,GlnK}}\right)\left(1+\frac{[\text{GLN}]}{K_\text{GLN,1}}\right)}\\
        &v_\text{UR,GlnB-U}=\frac{V_\text{UR}\frac{[\text{GlnB-U}]}{K_\text{m,GlnB-U}}}{\left(1+\frac{[\text{GlnB-U}]}{K_\text{m,GlnB-U}}+\frac{[\text{GlnK-U}]}{K_\text{m,GlnK-U}}\right)\left(1+\frac{K_\text{GLN,1}}{[\text{GLN}]}\right)}\\
        &v_\text{UT,GlnK}=\frac{V_\text{UT}\frac{[\text{GlnK}]}{K_\text{m,GlnK}}}{\left(1+\frac{[\text{GlnB}]}{K_\text{m,GlnB}}+\frac{[\text{GlnK}]}{K_\text{m,GlnK}}\right)\left(1+\frac{[\text{GLN}]}{K_\text{GLN,1}}\right)}\\
        &v_\text{UR,GlnK-U}=\frac{V_\text{UR}\frac{[\text{Glnk-U}]}{K_\text{m,Glnk-U}}}{\left(1+\frac{[\text{GlnB-U}]}{K_\text{m,GlnB-U}}+\frac{[\text{GlnK-U}]}{K_\text{m,GlnK-U}}\right)\left(1+\frac{K_\text{GLN,1}}{[\text{GLN}]}\right)}
    \end{align}
    \end{subequations}
    }
    Here $V_{*}$ and $K_\text{m,*}$ are the maximal rates and Michaelis constants of the (de)uridylylation reactions, and $K_\text{GLN,1}$ is the binding affinity of GLN to the allosteric site of UT/UR. The concentration of GlnB (and similarly GlnK) is obtained from the conservation relation: 
    {\small
    \begin{align}
        [\text{GlnB}]=[\text{GlnB}_{\text{total}}](t)-[\text{GlnB-U}]
    \end{align}
    }
    where the total concentration is a time-varying experimental measurement (Fig. \ref{drivers}). 
    
\paragraph*{\textbf{Sequestration of GlnK:}}
    We assume that GlnK-AmtB complex formation regulates the amount of GlnK accessible to other reactions, and is controlled by $\alpha$-KG following a Hill-type form~\cite{hwa}:
    {\small
    \begin{align}
        [\text{GlnK}]=[\text{GlnK}_\text{total}] \left(x+(1-x)\frac{[\alpha\text{-KG}]^{n_\text{UT}}}{K^{n_\text{UT}}_\text{$\alpha$-KG,1}+[\alpha\text{-KG}]^{n_\text{UT}}}\right)
        \label{hill}
    \end{align}
    }
where $[\text{GlnK}_\text{total}]$ and $[\text{GlnK}]$ are the total and accessible concentrations of GlnK, respectively; $K_\text{$\alpha$-KG,1}$ is the dissociation constant for $\alpha$-KG; $n_\text{UT}$ is the Hill-coefficient; and $x$ is the minimum concentration of accessible GlnK, which indicates the ratio of AmtB to GlnK. If $[\alpha\text{-KG}] \ll K_\text{$\alpha$-KG,1}$, then  $[\text{GlnK}]=x [\text{GlnK}_\text{total}]$, and the maximal amount of GlnK is sequestered; if $[\alpha\text{-KG}] \gg K_\text{$\alpha$-KG,1}$, then $[\text{GlnK}]\approx x[\text{GlnK}_\text{total}]$, so no GlnK is sequestered.
    
\paragraph*{\textbf{Equations for AT/AR:}}

    We split the activity of AT/AR into independent AT and AR reactions, both obeying Hill-type kinetics. We consider that GlnB-U/GlnK-U and GlnB/GlnK are pairwise competing for allosteric sites, while glutamine can bind independently to a third allosteric site \cite{jaggi,ninfa9,ninfa13}. The binding of GlnB and $\alpha$-KG are synergistic, and that of GlnB-U and $\alpha$-KG are antagonistic \cite{ninfa12} leading to the following activation parameters:
    {\small
    \begin{subequations}\label{effectiveK}
    \begin{align}
        &K_\text{GlnB}=\overline{K_\text{GlnB}}\left(1+\frac{[\alpha\text{-KG}]}{K_\text{$\alpha$-KG,2}}\right) \\
        &K_\text{GlnB-U}=\overline{K_\text{GlnB-U}}\left(1+\frac{K_\text{$\alpha$-KG,2}}{[\alpha\text{-KG}]}\right).
    \end{align}
    \end{subequations}
    }
Similar expressions are used for GlnK and GlnK-U. As above, we only describe GS-A monomers (without explicit GS-A complexes)~\cite{mutalik,Bruggeman,ninfa16}. We neglect retroactivity from GlnB and GlnK on $\alpha$-KG, since the latter is two orders of magnitude more abundant, and from AT/AR on GlnB and GlnK, since GlnK level is expected to be significantly higher \cite{ninfa9}. These assumptions result in the following rate equations (see Section S2 in Supporting Material for a detailed derivation):
    {\small
    \begin{subequations} \label{ATAR}
    \begin{align}
        v_\text{AT}&=\frac{V_\text{AT}[\text{GS}]^{n_\text{AT}}}{K_\text{m,GS}^{n_\text{AT}}+[\text{GS}]^{n_\text{AT}} } \left[ \frac{1}{\left(1+\frac{K_\text{GlnB}}{ [\text{GlnB}]}+\frac{[\text{GlnK}]}{K_\text{GlnK}}\frac{K_\text{GlnB}}{[\text{GlnB}]}\right)} \right. \notag\\ & \left.+\frac{1}{\left(1+\frac{K_\text{GlnK}}{[\text{GlnK}]}+\frac{[\text{GlnB}]}{K_\text{GlnB}}\frac{K_\text{GlnK}}{[\text{GlnK}]}\right)} \right]\frac{1}{1+\frac{K_\text{GLN,2} }{[\text{GLN}]}} \\
        v_\text{AR}&=\frac{V_\text{AT}[\text{GS-A}]^{n_\text{AT}}}{K_\text{m,GS-A}^{n_\text{AT}}+[\text{GS-A}]^{n_\text{AT}}}\left[\frac{1}{\left(1+\frac{K_\text{GlnB-U}}{[\text{GlnB-U}]}+\frac{[\text{GlnK-U}]}{K_\text{GlnK-U}}\frac{K_\text{GlnB-U}}{[\text{GlnB-U}]}\right)}\right.
        \notag\\&\left.+\frac{1}{\left(1+\frac{K_\text{GlnK-U}}{[\text{GlnK-U}]}+\frac{[\text{GlnB-U}]}{K_\text{GlnB-U}}\frac{K_\text{GlnK-U}}{[\text{GlnK-U}]}\right)} \right] \frac{1}{1+\frac{[\text{GLN}]}{K_\text{GLN,2}}}
    \end{align}
    \end{subequations}
    }
where $V_{*}$ and $K_{m,*}$ are the maximal enzyme rate and Michaelis constants of (de)adenylylation, $K_{*}$ are the binding affinities of GlnB(-U), GlnK(-U) and GLN to the allosteric sites of AT/AR and $n_\text{AT}$ is the Hill-coefficient.
A summary of these interactions is shown in Figure~\ref{enzUTUR}. 

Simulations were performed in Matlab. An SBML format of the model is included in the Supporting Material where we refer the reader for further details.

\subsection*{Model parameters: fitting and sensitivity analysis}

\begin{table}[!b]
    \centering
    \caption{\small{Model parameters}}
    \vspace{2mm}
    \label{modelpars}
    \resizebox{1\textwidth}{!}{%
    \begin{tabular}{l | lcc | lcc}
    &\multicolumn{3}{c|}{\textbf{UT/UR}} & \multicolumn{3}{c}{\textbf{AT/AR}} \\
   &  & Literature & Fitted &  & Literature & Fitted \\
    \hline
   \multirow{2}{6em}{Maximum enzyme rates} &$V_\text{UT}$ & - & 119.7 $\mu$M/min & $V_\text{AT}$ & - & 976.9 $\mu$M/min \\
    &$V_\text{UR}$ & - & 14.7 $\mu$M/min & $V_\text{AR}$ & - & 473.4 $\mu$M/min \\
    \hline
    \multirow{4}{6em}{Michaelis-Menten constants} &$K_\text{m,GlnB}$ & 3 $\mu$M \cite{ninfa3} & - & $K_\text{m,GS}$ & - & 133.8 $\mu$M \\
    &$K_\text{m,GlnB-U}$ & 2 $\mu$M \cite{ninfa3} & - & $K_\text{m,GS-A}$ & - & 56.1 $\mu$M \\
    &$K_\text{m,GlnK}$ & - & 9.9 $\mu$M &  &  &   \\
    &$K_\text{m,GlnK-U}$ & - & 59.8 $\mu$M &  & &  \\
    \hline
    \multirow{6}{5em}{Dissociation constants} &$K_\text{GLN,1}$ & 90 $\mu$M \cite{ninfa3} & - & $\overline{K_\text{GlnB}}$ & 15.0 $\mu$M \cite{ninfa9} & -  \\
   & $K_\text{$\alpha$-KG,1}$ & - & 3.0 mM & $\overline{K_\text{GlnB-U}}$ & 0.25 $\mu$M \cite{ninfa9}   & - \\
   &  &  &  & $\overline{K_\text{GlnK}}$ & -  & 8.3 $\mu$M \\
   &  &  &  & $\overline{K_\text{GlnK-U}}$ & - & 9.7 $\mu$M \\
   &  &  &  & $K_\text{GLN,2}$ & - & 1.4 mM \\
   &  &  &  & $K_\text{$\alpha$-KG,2}$ & - & 0.89 mM \\  
     \hline
    Hill coefficients & $n_\text{UT}$ & 3 \cite{hwa}  & -  & $n_\text{AT}$ & - & 2\\
    \hline
    Fractional sequestration& $x$ & - & 0.4 & & &     
    \end{tabular}%
    }
\end{table}

    The parameters were taken from the literature whenever possible (6 out of 21) and the rest (15 out of 21) were fitted to the 36 time points of the WT data (Table~\ref{modelpars}). 
    
To assess the robustness of the model and to evaluate the relevance of the parameters to be fitted, we performed a global sensitivity analysis using the eFAST algorithm~\citep{saltelli}. This technique considers uncertainty in combinations of parameters and quantifies the deviation they introduce over the whole time series. See Section S3 in the the Supporting Material for details of the method and its results.

   To fit the parameters of the model to the measured time series, we used the Squeeze-and-Breathe evolutionary Monte Carlo optimisation algorithm~\cite{beguerisse}. This algorithm minimises a cost function $J$, which is the sum-of-squares deviation of the fractional PTM levels from the measurements, weighted by the relative errors. Mathematically,
    \begin{align}\label{costfun}
        J(p)=\sum_i\sum_j\norm{\frac{M^\text{mod}}{M^\text{total}}-\frac{F^\text{mod}(p)}{M^\text{total}}}\left(\frac{M^\text{total}}{M^\text{error}}\right),
    \end{align}
where $p$ denotes the vector of parameters; $M$ and $F$ denote measured and fitted, respectively; and the superscripts $mod$, $total$ and $error$ denote total, modified and error, respectively. The parameter values are presented in Table~\ref{modelpars}, rounded to the nearest decimal found in the confidence interval given by the distribution of optimised parameters obtained with the Squeeze-and-Breathe method (see Section S4 in the Supporting Material for details).

\section*{Results}

\subsection*{\textit{In vivo} measurements of \textit{E.~coli} temporal responses to nitrogen starvation and upshift}
    
\begin{figure}[t!]
     \centering
     \includegraphics[width=65mm]{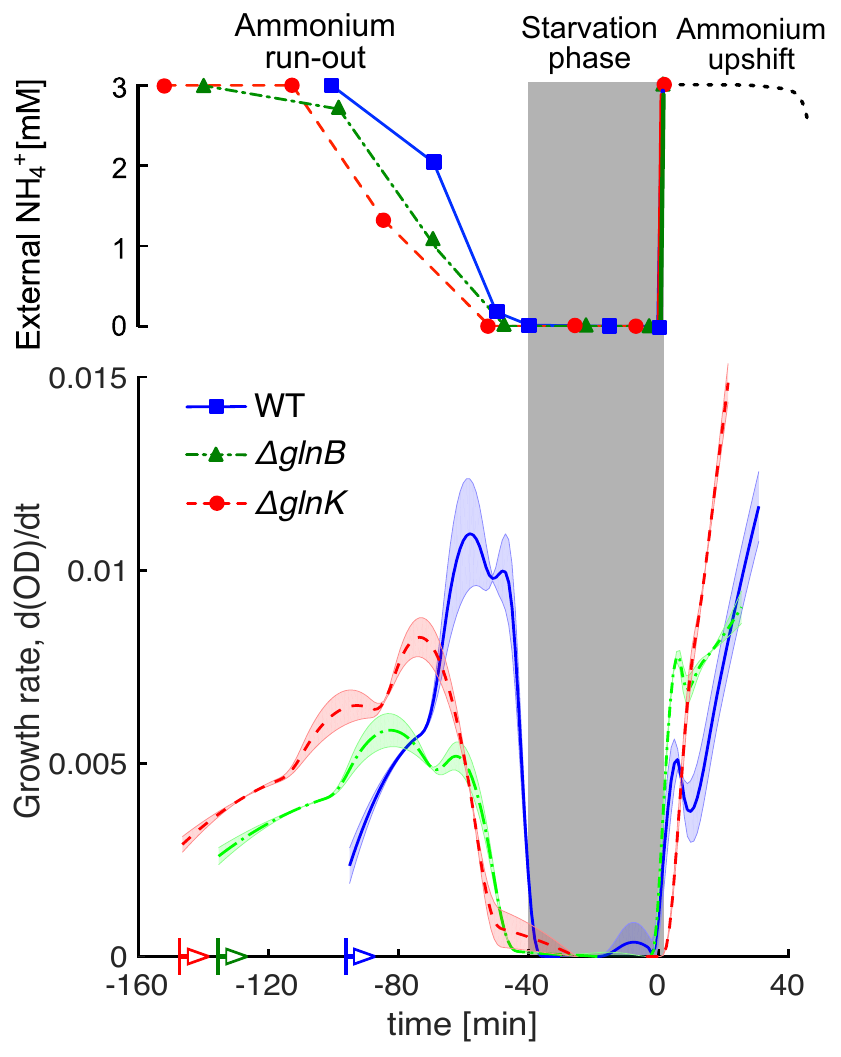}
     \caption{Experimental time series of growth rate of the WT and the two mutants ($\Delta$\textit{glnB} and $\Delta$\textit{glnK}) during ammonium run-out followed by a step change of ammonium. During the starvation phase (marked by the grey shaded area), the growth is halted. The time series of the three strains were aligned such that the onset of starvation coincides. The three strains were introduced to the medium at different times, as indicated by the coloured arrows. The lines and symbols represent averages and the coloured bands represent $\pm$s.e (n=3). The growth rate was computed using a Savitzky-Golay filter applied to the experimental measurements of the optical density (OD) over the course of the experiment.}
     \label{growth}
\end{figure}
    
We measured concentrations of relevant metabolites ($\alpha$-KG and GLN) and active/inactive PTM states of the proteins GlnB, GlnK and GS in response to dynamic variations in external ammonium (see Materials and Methods). We use this novel experimental approach to measure PTM states~\cite{Schumacher,komorowski} in order to link metabolic status and regulatory mechanisms.

Three \textit{E. coli} cultures (a WT strain and two isogenic strains with deleted \textit{glnK} or \textit{glnB}) were inoculated into minimal media with a (limiting) concentration of 3 mM ammonium (see Materials and Methods). The growing cultures depleted the external nitrogen levels during the run-out period eventually leading to the starvation period measured from the point when growth stopped and extending for 40 minutes (Fig.~\ref{growth}). The starvation period was followed by a sudden 3 mM ammonium upshift to investigate the shock response (0 min to 40 min). We sampled the cultures at specific external ammonium concentrations during run-out and at specific time points during starvation and upshift, and measured the metabolite and protein concentrations. The corresponding time courses are plotted on Fig. \ref{drivers}.

\subsection*{\textit{In vivo} regulation of GS by GlnK}
     
\begin{figure*}[t!]
        \centering
        \includegraphics[width=1\textwidth]{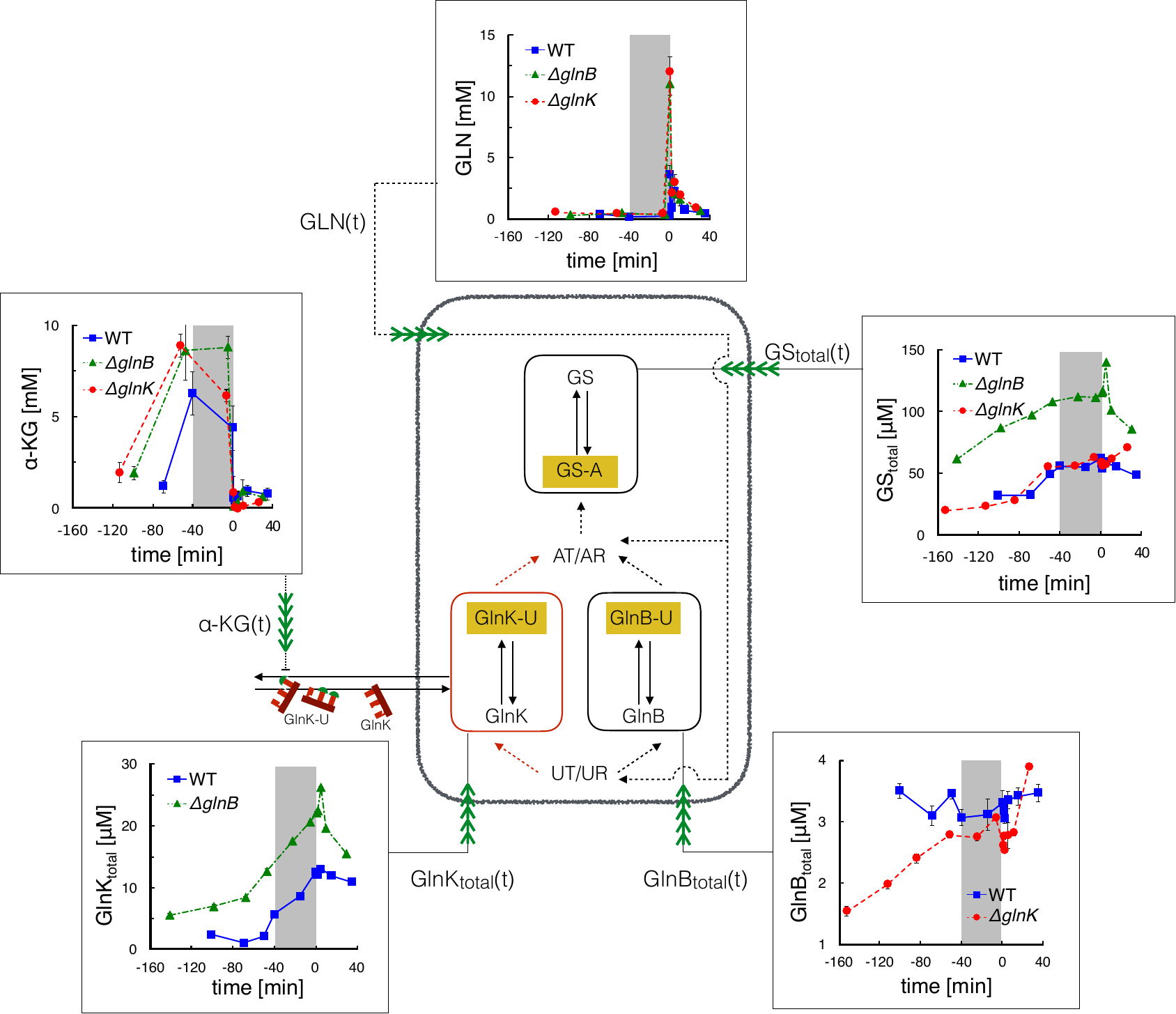}
        \caption{Experimental time series of total monomeric protein (GlnB, GlnK and GS) and metabolite (glutamine and $\alpha$-KG) concentrations for the WT (blue), $\Delta$\textit{glnB} (green) and $\Delta$\textit{glnK} (red) strains. Symbols and error bars represent average protein or metabolite concentrations $\pm$s.e (n=3). The grey shaded area marks the starvation period. The lines are guides to the eye. This data was used to drive the model (pictured) of the PTM of the three proteins (highlighted in orange). Time-course measurements of the modified proteins GS-A, GlnB-U and GlnK-U were also taken experimentally and are presented in Fig.~\ref{fig_rev}.
        }
        \label{drivers}
    \end{figure*}

First, we investigated whether our experiments provide evidence for GlnK controlling the level of GS expression. We found no discernible difference in total GS level between $\Delta$\textit{glnK} and WT (Fig. \ref{drivers}), indicating that GlnK is redundant to regulate GS expression in the WT. In addition, we found that GlnK and GS were two-fold over-expressed in $\Delta$\textit{glnB} (Fig.~\ref{drivers}). This observation provides an \textit{in vivo} quantification of the reduced regulatory effect of GlnK on GS expression, which has been previously evidenced in genetic studies~\cite{ninfa7}. 
    
Next, we asked whether GlnK plays a relevant role as a regulator of GS adenylylation. The response of the $\Delta$\textit{glnB} strain to external ammonium variations was consistent with GlnK substituting for GlnB in its role to control AT/AR \cite{ninfa8}. Although the growth of $\Delta$\textit{glnB} was slower than the WT during ammonium run-out, it was comparable to $\Delta$\textit{glnK} (Fig.~\ref{growth}), indicating that, relatively speaking, ammonium uptake was not limiting. We further found that GS was abruptly adenylylated upon nitrogen upshift in all strains (Figs. ~\ref{fig_rev}c,e,g), including the $\Delta$\textit{glnB} mutant, despite higher GS levels in the latter (Fig.~\ref{drivers}). This suggests that, after starvation, GlnK is a potent activator of GS adenylylation. In contrast, GS was not fully deadenylylated in $\Delta$\textit{glnB} (Fig.  ~\ref{fig_rev}e and Fig. S6a in the Supporting Material) under ammonium rich conditions (unlike in WT and in $\Delta$\textit{glnK}, see Fig.~\ref{fig_rev}c,e), showing that GlnK might be less efficient in activating the AR activity. Indeed, GlnK was almost fully uridylylated ($>$95\%) during starvation (see Fig. S6c in the Supporting Material), consistent with the dominant GlnK species under starvation being the fully uridylylated GlnK trimer, GlnK\textsubscript{3}-U\textsubscript{3}, which is known to be a poor activator of deadenylylation \textit{in vitro}~\cite{Heeswijk4}. 

Unlike in $\Delta$\textit{glnB}, we observed no evidence of GlnK regulated GS adenylylation in the WT during ammonium run-out and starvation, the GS-A levels being similar between the WT and $\Delta$\textit{glnK} (Fig.~\ref{fig_rev}c,g). Despite this purported redundancy, the expression and uridylylation of GlnK are already induced when ammonium levels begin to decrease~\cite{Heeswijk2} (Fig. \ref{drivers} and \ref{fig_rev}b). Since GlnB expression is constitutive, we find that during starvation the concentration of GlnK (12.5 $\mu$M) largely exceeds that of GlnB (3.3 $\mu$M) (Fig.~\ref{drivers}) and similarly GlnK-U (11 $\mu$M) largely exceeds GlnB-U (2.9 $\mu$M) (Fig. \ref{fig_rev}a,b). Thus intracellular glutamine sensing by UT/UR operates before ammonium concentration becomes suboptimal for growth. 
    
Taken together, these findings suggest a role for GlnK to prepare for a buffered return to subsequent normal nitrogen conditions after starvation, by regulating the inactivation of GS. Our measurements show that the $\Delta$\textit{glnK} strain grows at a higher rate after ammonium upshift (Fig. ~\ref{growth}). We hypothesise that this difference is due to additional ammonium being imported via the membrane-bound ammonium channel AmtB, which accumulates during nitrogen starvation and whose activity becomes unimpeded at low levels of GlnK.

To understand further the interaction between GlnB and GlnK during nitrogen run-out, starvation and upshift, we constructed a mathematical model motivated by our \textit{in vivo} dynamic measurements of protein (total and PTM) and metabolite concentrations. To date, the use of directly determined PTM data in models of regulation of GlnB and GlnK has been lacking, yet such modification directly determines their functionality. We now present an inclusive modelling approach to link GlnB and GlnK activity via PTMs to metabolic status and control of nitrogen assimilation processes.

\subsection*{Mathematical modelling of the enzymatic network of nitrogen signalling and regulation}
\label{sec:model}
    
In view of the experimental data, we built an ordinary differential equation (ODE) model to describe the dynamics of PTM states, i.e., the (de)uridylylation of GlnB and GlnK and the (de)adenylylation of GS in response to temporal inputs (drivers). An ODE model was adopted due to the large copy numbers observed---the smallest being GlnB with an average of $\sim10^3$~molecules per cell. We based our equations on the enzyme mechanisms reported in the literature with the aim of a parsimonious model, only including terms with an observable effect in our experiments. The resulting model remains mechanistic with a moderate number of parameters, facilitating direct comparison to experiments and parameter fitting.

The terms of the ODE model are derived from the biochemical details of competition for the active and allosteric sites, which is inherent to the uridylylation and adenylylation reactions, and from the formation of complexes between GlnK and AmtB. In addition, we use five measured input drives to isolate the responses pertaining to transcription and other parts of the \textit{E. coli} metabolism. These include concentrations of metabolites GLN and $\alpha$-KG, which act as enzyme effectors and reflect the metabolic state of the cell, and the total concentrations of GlnB, GlnK and GS, which reflect the relevant transcriptional responses of the cell (Fig.~\ref{drivers}). We further assume that key cofactors (e.g., ATP, NADPH) remain at homeostasis under our experimental conditions~\cite{doucette}, so that their concentrations can be approximated to be constant and can thus be absorbed into the kinetic parameters of the equations.

The model parameters include: binding affinities ($K_\text{d}$) of various effectors to active and allosteric sites of the enzymes; Michaelis constants ($K_\text{m}$) and maximum enzyme rates ($V_\text{max}$); and parameters representing AmtB-GlnK complex formation (Table \ref{modelpars}). Whenever possible we used parameters from the biochemical literature, and we fitted the rest to the WT experimental time-courses using the Squeeze-and-Breathe algorithm~\cite{beguerisse}. To confirm that the parameters capture meaningfully the experimental observations, we also performed a global sensitivity analysis over the time series~\citep{saltelli}. For details, see Materials and Methods and the Supporting Material. 

To make predictions for the $\Delta$\textit{glnB} and $\Delta$\textit{glnK} mutants, we set the corresponding reaction rates to zero, fixing all other parameters at values fitted to the WT time-course data. We then generate predicted time series for the PTM protein states under the experimentally measured drives. These predictions are compared to the independently obtained measurements.

\subsection*{Modelling the mechanism of GlnB and GlnK uridylylation}

\begin{figure}[t!]
        \centering
        \includegraphics[width= 0.8\textwidth]{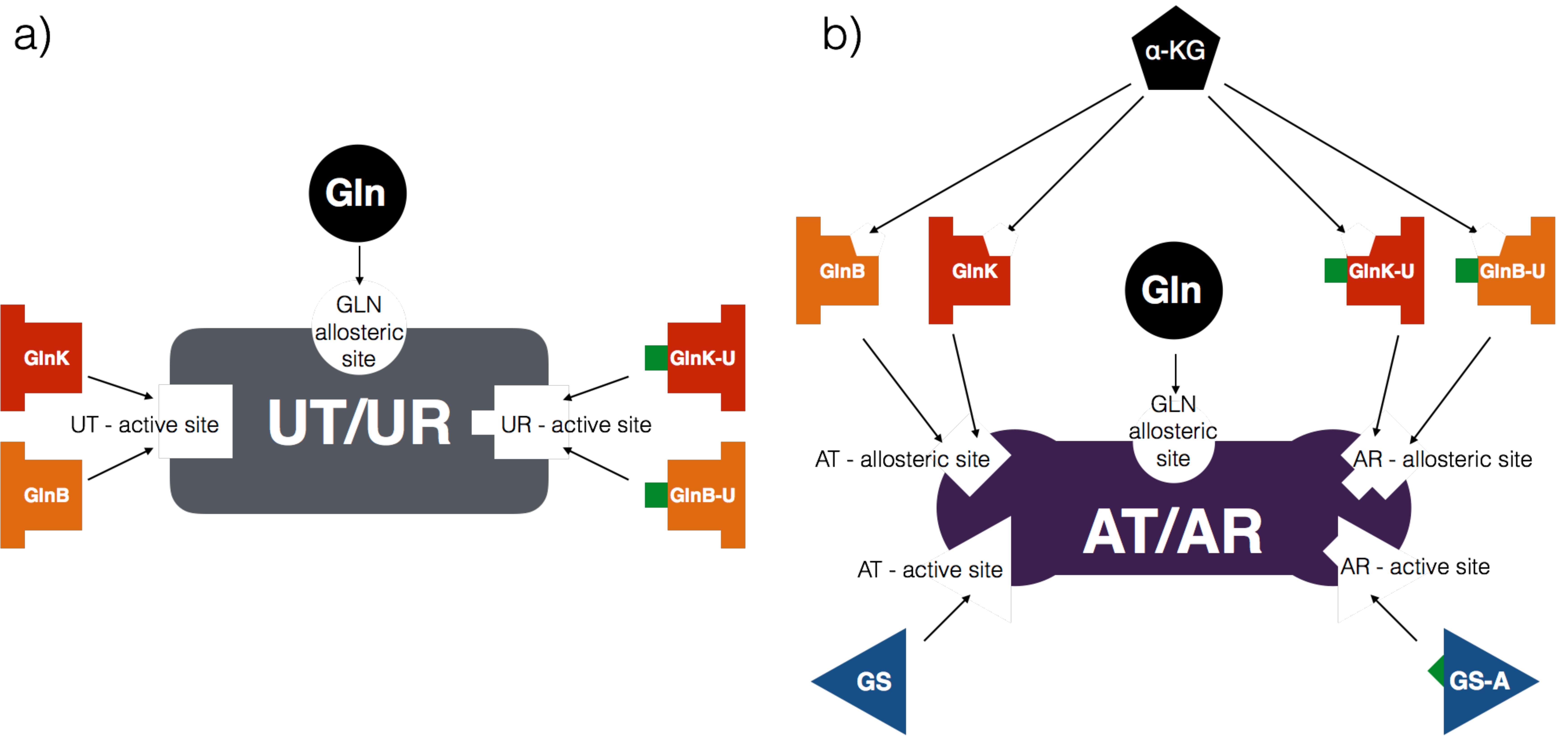}
        \caption{Schematic illustration of the structure of UT/UR and AT/AR. a) UT/UR has two distinct active sites: one binds GlnB-U and GlnK-U, whereas the other binds the unmodified GlnB and GlnK species. Hence GlnB and GlnK are competing substrates for UT/UR. In addition, UT/UR is allosterically regulated by glutamine, which activates the UR and inhibits the UT activity. b) AT/AR has two separate active sites: one binds GS, the other GS-A. It also has three distinct allosteric sites: the first binds GlnB-U and GlnK-U and induces AR excitatory and AT inhibitory responses; the second binds GlnB and GlnK and induces AT excitatory and AR inhibitory responses; the third binds GLN and modulates the rate of AT and AR responses. In addition, $\alpha$-KG can bind to both the modified and unmodified GlnB/GlnK proteins, affecting their interaction with AT/AR. }
        \label{enzUTUR}
    \end{figure}

The bifunctional enzyme UT/UR has two independent active sites catalysing uridylylation (UT) and deuridylylation (UR), which are highly specific for GlnB and GlnB-U, respectively~\cite{Zhang}; it also has an allosteric site for glutamine (Fig. \ref{enzUTUR}a). Due to the similarity between GlnB and GlnK, we assume similar specificity of UT/UR active sites to GlnK/GlnK-U, yet with different binding affinities. Hence we model GlnB and GlnK uridylylation as two monocycles, GlnB-UT/UR and GlnK-UT/UR, which operate in parallel but not independently, since the total amount of enzyme is distributed between them at any given time. We further consider that the binding of glutamine inhibits the uridylylation of GlnB/GlnK and activates the deuridylylation of GlnB-U/GlnK-U~\cite{ninfa1} (Eq. \ref{UTUR}, Materials and Methods). 

As shown in Fig.~\ref{fig_rev}a, our model captures accurately the excitatory and inhibitory effects in the measured GlnB uridylylation under different conditions in the WT without the need for modelling ternary complexes explicitly as in the more complex model in Ref.~\cite{straube} (Fig. S5 in the Supporting Material). The parameters fitted to the WT data show that the affinities of GlnK and GlnK-U to AT/AR ($K_\text{m}=9.9 \mu$M and 59.8$\mu$M, respectively) are significantly lower than those of GlnB and GlnB-U ($K_\text{m}=3 \mu$M and 2$\mu$M, respectively), which are known experimentally \cite{ninfa3}. Hence GlnK only affects nitrogen assimilation when its concentration is several times higher than GlnB, in agreement with previous \textit{in vitro} findings~\cite{ninfa7}. 

Although this model allows us to fit well the overall WT time measurements, it cannot capture the GlnK-U level in response to the ammonium upshift after starvation (Fig.~\ref{fig_rev}a).
Moreover, the model over-predicts the level of GlnK-U in the $\Delta$\textit{glnB} knock-out (Fig.~\ref{fig_rev}b). 
In the next section we discuss an additional mechanism, which is necessary to make accurate predictions of the observed GlnK-U levels.

\subsection*{Sequestration of GlnK is necessary to predict uridylylation levels} 

    To accurately describe GlnK uridylylation dynamics, we take into account the interaction of GlnK with AmtB. The membrane-bound channel protein AmtB facilitates ammonium active import to the cell. AmtB channels are blocked when AmtB forms a complex with trimers of deuridylylated GlnK above 20-50 $\mu$M external ammonium concentrations~\cite{merrick1}. Hence AmtB-facilitated transport is negligible in our experiments before complete run-out. However, the formation of complexes between AmtB and GlnK does affect the dynamical responses, since it provides  a mechanism for the \textit{sequestration} of GlnK and GlnK-U at the membrane~\cite{merrick4}. Note that while fully uridylylated trimers cannot bind AmtB \cite{merrick3}, our measurements show that the uridylylation level of GlnK is around 30\% after starvation (see Fig. S6c in the Supporting Material) suggesting that the GlnK trimers will be mostly mono-uridylylated, i.e. one uridylylated GlnK per GlnK trimer, on average. In our model, we assume that both fully deuridylylated GlnK trimers and partially uridylylated GlnK trimers may be functionally sequestered to the membrane. However, unlike the fully deuridylyated GlnK trimers,  partially uridylylated GlnK trimers do not block AmtB channel activity.  
        
    We model AmtB-GlnK complex formation by allowing a proportion of the total GlnK to be sequestered with a Hill-function dependence (Eq.~\ref{hill}, Materials and Methods) on the concentration of $\alpha$-KG~\cite{merrick1}. Note that, since high levels of $\alpha$-KG level promote the dissociation of the AmtB-GlnK complex, sequestration only occurs during external ammonium abundance, i.e., when the $\alpha$-KG levels are low (Fig. \ref{drivers}).
    
\begin{figure}[t!]
	\centering
    \includegraphics[width=0.95\textwidth]{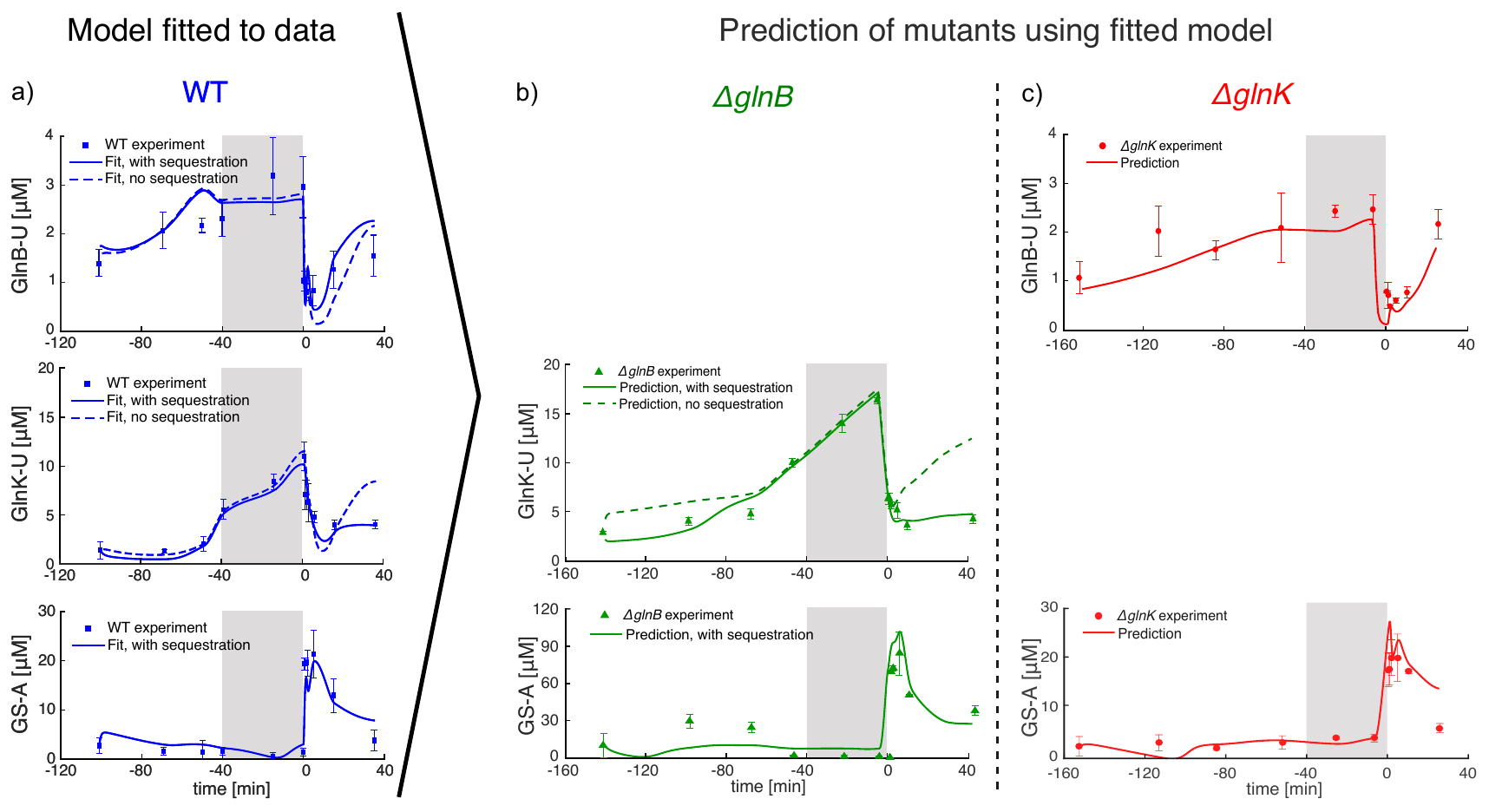}
    \caption{Time responses of PTM protein levels in the WT and two knock-out mutants. a) Time responses of GlnB uridylylation, GlnK uridylylation and GS adenylylation in the WT strain (experimental measurements and fitted model, with and without sequestration).  b-c) Experimental measurements of GlnB uridylylation, GlnK uridylylation and GS adenylylation for knock-outs: b) $\Delta$\textit{glnB} (green) and c) $\Delta$\textit{glnK} (red), compared in both cases to the dynamics predicted by our model fitted only to the WT data. The symbols and error bars represent measured average protein concentrations $\pm$s.e (n=3). The lines correspond to fits and prediction of our model, as indicated. The grey shaded area marks the starvation period. The model with sequestration provides a significantly better fit according to information criteria for model selection (see text).
}
    \label{fig_rev}
\end{figure}

The model with sequestration fitted to the WT data is able to capture the GlnK-U levels after nitrogen upshift (Fig.~\ref{fig_rev}a). To justify the necessity of the two additional parameters ($ K_{\alpha\text{-KG},1} $, $ x $) involved in GlnK sequestration, we use the fitted model to predict the uridylylation states of GlnK and GlnB in the $\Delta$\textit{glnK} and $\Delta$\textit{glnB} knock-outs, and compare them to the measured time-courses (Fig.~\ref{fig_rev}b,c). 
This approach has been used before to develop systems biology models~\cite{link,doucette2}. Our predictions agree with the measurements: without sequestration, the model overpredicts the levels of GlnK-U before starvation (time before -40 min) and after nitrogen upshift (time after 0 min) in the $\Delta$\textit{glnB} mutant (Fig.~\ref{fig_rev}d). To quantify this, we compute the error of both models (with and without sequestration) against the data~\eqref{costfun}, and rank both models according to two information theoretic criteria (AICc, BIC), which penalise the additional parameter complexity against any improvement in the error of the model. The extended model with sequestration is clearly selected by both information criteria (see Table S2 in the Supporting Material) lending further basis for the need for a GlnK sequestration term during nitrogen excess. 

The higher growth rate observed in the $\Delta$\textit{glnK} strain after starvation (Fig. \ref{growth}) is also consistent with GlnK sequestration. Since \textit{amtB} is still present in this strain, AmtB is expected to accumulate during nitrogen starvation, which provides unrestricted ammonium import and higher growth rate in the short term.

\subsection*{GlnK can regulate AT/AR in the absence of GlnB}

    Our results so far indicate that GlnK not only competes with GlnB as a substrate or UT/UR, but also provides a link between active membrane transport and carbon metabolism through the interaction with $\alpha$-KG. To further test this mechanism, we asked: is GlnK sufficient on its own to regulate the adenylylation enzyme AT/AR in the absence of GlnB? If so, one would expect that the predicted GlnK-U levels in the $\Delta$\textit{glnB} strain should account correctly for GS-A levels and, hence, for the activity of GS in nitrogen assimilation.
    
    To test this hypothesis, we extended our model to include GS adenylylation. Similarly to UT/UR, the two antagonistic activities of AT/AR are modelled as independent uni-directional reactions, justified by structural studies showing the two active sites to be well separated~\cite{jaggi}, making interaction between the AT and AR domains less likely (Fig.~\ref{enzUTUR}b). 
     
We model different facets of AT/AR regulation as found in the literature (Fig.~\ref{enzUTUR}b). Firstly, AT and AR activities are inhibited by the products, GS-A and GS, respectively \cite{ninfa9}. Secondly, in the absence of allosteric effectors AT/AR has no observable activity \cite{ninfa9}. From this state, the rates of AT and AR actitivites are regulated by five allosteric effectors, which no not affect the binding rates of substrates \cite{ninfa13}. In particular, GLN, GlnB and GlnB-U may bind to three distinct sites~\cite{jaggi,ninfa9,ninfa13}. As above, we assume that GlnK(-U) has the same specificity as GlnB(-U) but different affinity, so that GlnB and GlnK compete for one allosteric site, and GlnB-U and GlnK-U compete for the other. Finally, to account for intramolecular signalling \cite{jaggi}, we allow the activation parameters of GlnB, GlnB-U, GlnK and GlnK-U to depend on $\alpha$-KG (Eqs. \ref{effectiveK}, Materials and Methods). We encapsulate these terms into the reaction schemes (Eq.~\ref{ATAR}), by generalising the classical allosteric activation model~\cite{cornish} for two competing effectors (see Section S2 in the Supporting Material for details).        
                
    The additional equation for GS-A was fitted to the WT data and captures well the observed variability (Fig. \ref{fig_rev}a). To avoid overfitting, we keep fixed all parameters related to UT/UR and sequestration, and fix the insensitive activation parameters $\overline{K_\text{GlnB}}$ and $\overline{K_\text{GlnB-U}}$ to the values reported by \cite{ninfa9} (Table \ref{modelpars}, Materials and Methods and Section S3 in the Supporting Material). In contrast, the activation parameters $\overline{K_\text{GlnK}}$ and $\overline{K_\text{GlnK-U}}$ are relatively insensitive in the WT, yet essential to predict GS-A levels in the $\Delta$\textit{glnB} strain. The fitted activation constant $K_\text{GLN,2}$ for AT/AR is in the range reported in Ref.~\cite{ninfa9} and the Hill coefficient $n_\text{AT}$ is found to be as reported in Ref.~\cite{mutalik}. Note also that the activation constants of GlnK and GlnK-U for AT/AR are much higher than those of GlnB and GlnB-U, consistent with their observed lower affinity~\cite{ninfa7}.  
	
    To assess the model fitted to the WT data, we predict the level of GS-A for the two knock-outs in response to the measured drives. The predicted responses agree with the measured GS-A levels in both strains (Fig.~\ref{fig_rev}b,c). The GlnB-deficient strain ($\Delta$\textit{glnB}) is able to regulate GS activity in response to external changes in ammonium. The prediction for the $\Delta$\textit{glnB} strain underestimates the GS-A level before starvation---we consider the possible source for this discrepancy in the Discussion. Given that GS regulation depends on the fine balance between GlnB/GlnB-U and GlnK/GlnK-U together with the level of GLN and $\alpha$-KG, our model demonstrates that GlnK can regulate GS under varying ammonium conditions.

\section*{Discussion}	

    We have investigated the role of GlnK in the dynamic regulation of ammonium assimilation from a systems perspective, and have argued that paradoxical observations about GlnK stem from the subtle interplay of regulatory mechanisms on two levels: transcriptional and post-translational. To decouple both levels of regulation, we collected concurrent \textit{in vivo} time-courses of metabolite, total protein and PTM protein concentrations in response to time-varying external ammonium levels. We analysed the data in conjunction with a mechanistic model derived from biochemical principles. The combined analysis allowed us to fit the WT dynamics of PTMs, and to used the fitted model to predict the dynamic responses of experimental data from genetic knock-outs, shedding light on experimental ambiguities about the regulatory role of GlnK. 
    
    We found that competition between GlnB and GlnK for the bifunctional enzyme UT/UR together with GlnK sequestration by AmtB describe the uridylylation responses in the WT and can predict the response of two mutants under dynamic ammonium perturbations. AmtB-GlnK complex formation not only blocks ammonium import~\cite{merrick1}, but also rapidly sequesters GlnK, whereas the less abundant GlnK limits ammonium assimilation by facilitating more rapid deuridylylation of GlnK-U (Fig.~\ref{fig_rev}) and thus higher GS-A levels. Thus GlnK dampens the assimilatory response to ammonium upshift, perhaps to avoid excess ammonium uptake causing the depletion of the glutamate pool~\cite{wingreen,kustu}. The fact that sequestration is regulated by $\alpha$-KG (a product of the carbon metabolism), suggests that nitrogen shock responses could also help to avoid untoward impacts on carbon status.
 
    Our experiments showed that during ammonium rich conditions, GlnK is present at low level and does not influence GS production and adenylylation, which is instead regulated by the more potent GlnB. However, due to the induction of \textit{glnK} during ammonium run-out, GlnK outnumbers GlnB four-fold during starvation (Fig. \ref{drivers}). To examine the capability of GlnK for AT/AR regulation after ammonium upshift, we developed a mathematical model of GS adenylylation based on allosteric competition between GlnB and GlnK for AT/AR. The model predicts accurately the abrupt increase in GS-A levels after ammonium upshift in both the $\Delta$\textit{glnB} and $\Delta$\textit{glnK} strains (Fig.~\ref{fig_rev}) showing that both GlnB and GlnK are independently sufficient for effective AT/AR regulation. Our model slightly underestimates the GS-A levels before starvation in the $\Delta$\textit{glnB} strain. Since the model was fitted to the WT, where both GlnB and GlnK are present, this mismatch may indicate unaccounted-for interactions which are disrupted when deleting the \textit{glnB} gene, e.g., GlnB and GlnK forming heterotrimers~\cite{Forchhammer,Heeswijk4,Jiang:2009}. 
 
\begin{figure}[t!]
	    \centering
	   \includegraphics[width= \textwidth]{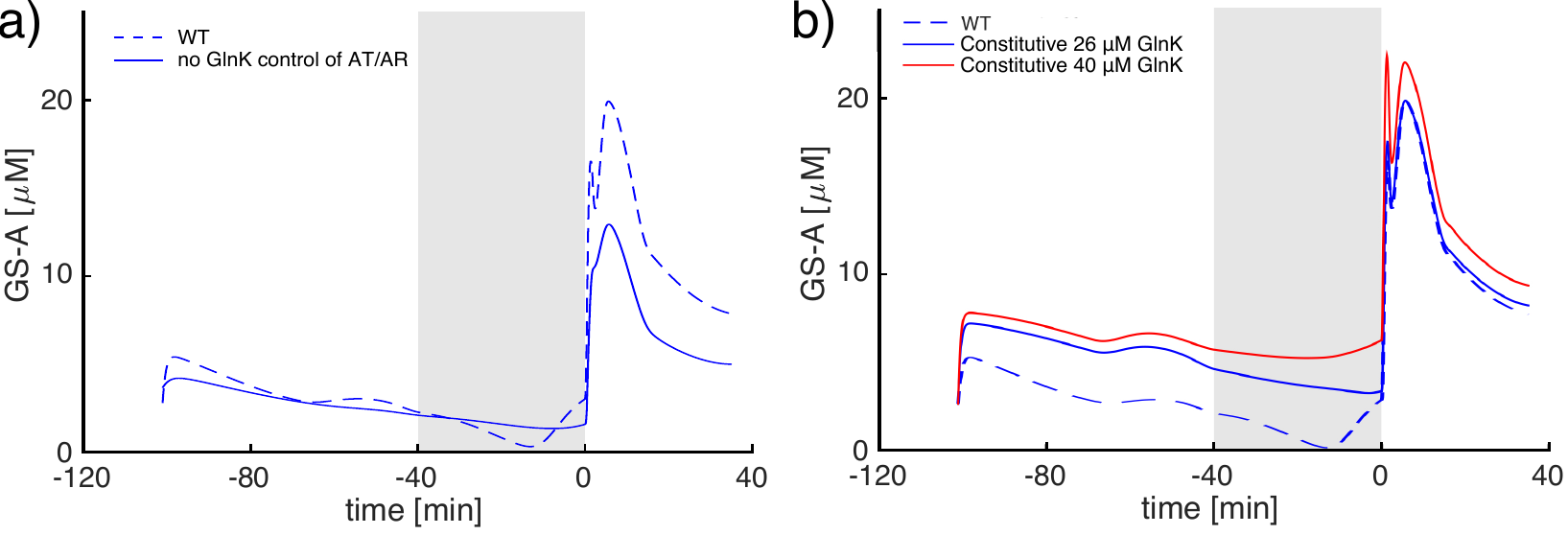}
	    \caption{Predictions of the model for the GS-A levels of two `computational strains'. a) In the first `strain', AT/AR is regulated only by GlnB but not by GlnK, yet GlnK still regulates AmtB and competes with GlnB for UT/UR. Before starvation, the levels of GS-A are not affected by GlnK due to low GlnK abundance. If GlnK is assumed not to regulate AT/AR, the level of GS-A  after ammonium upshift is reduced, implying increased ammonium assimilation. Hence the high level of GlnK induced during starvation in the WT acts as a buffer to prevent overshoots in ammonium assimilation by regulating GS activity. b) In the second `strain', the expression of \textit{glnK} (and conjointly of \textit{amtB}) is constitutive (unregulated). In order to avoid the overshoot due to ammonium upshift , a high level of GlnK is necessary before starvation. This over-abundance of GlnK pre-starvation leads to higher GS-A levels, and hence suboptimal ammonium uptake. In each figure, the grey shaded area indicates the ammonium starvation period; the dashed line shows the original model for the WT data (same as in Fig.~\ref{fig_rev}); the solid lines are predictions of the model for the respective `computational strains'.}
	    \label{predseq}
	\end{figure}

   To check the relevance of GlnK regulation of AT/AR, we used our model to simulate two `in silico' cellular scenarios. Firstly, we considered our model for a `computational strain' where GlnK does not interact with AT/AR (by setting $ \overline{K_\text{GlnK}} \gg [\text{GlnK}]$ and $\overline{K_\text{GlnK-U}} \gg [\text{GlnK-U}]$, such that the terms involving $ \text{GlnK} $ and $ \text{GlnK-U} $ drop out in Eq. \ref{ATAR}), and hence AT/AR is regulated only by GlnB. In this scenario where  the regulatory role of GlnK is removed, the predicted GS-A levels in response to ammonium upshift after starvation are reduced (Fig. \ref{predseq}a), thus leading to an increase in GS activity and larger ammonium assimilation. This is consistent with the proposed role of GlnK as a buffer to ammonium shock, through the adenylylation (inactivation) of GS. Noting that the level of GS-A does not change during run-out and starvation, we also used our model to predict the response of another `computational strain' (Fig. \ref{predseq}) where the expression of \textit{glnK} is constitutive and unregulated (by fixing $ [\text{GlnK}_\text{total}] $ as a constant). In this scenario, if GlnK abundance is kept fixed at the late-starved level (26 $\mu$M), GS-A levels remain elevated during run-out and starvation. This simulation supports the notion that GlnK must be kept down-regulated during nitrogen abundance, since early \textit{glnK} expression would lead to reduced ammonium assimilation due to increased GS-A levels. During ammonium run-out and starvation, \textit{amtB} (and conjointly \textit{glnK}, being on the same operon) is expressed so as to increase ammonium uptake via AmtB channels.  The resulting higher levels of GlnK induced by starvation serve also to control potential overshoots in ammonium levels following starvation not only by blocking AmtB channels, but also by directly reducing GS activity through adenylylation.
    
    Beyond nitrogen regulation, our work exemplifies the need for \textit{in vivo} experimental time series data in developing predictive models. Time-course inputs together with classic genetic knock-outs provide tools to gain insight into the dynamic aspects of regulation in signalling pathways. Although large-scale biochemical models can be powerful, their applicability is many times hampered due to the large number of unknown parameters they contain, which need to be fitted to scarce data. Here we derived a small set of mathematical modules commensurate with the observed dynamics, and justified their relevance by a global parameter sensitivity analysis. We used a recently developed parameter fitting algorithm~\cite{beguerisse}, which is especially appropriate for time-courses, to fit the WT data, and we then used the fitted model to predict out-of-sample dynamical observations from the genetic knock-outs.
    
Although we have focussed here on \textit{E. coli}, with the dual GlnB/GlnK proteins, we expect nitrogen assimilation and transport to be more strongly coupled in bacteria containing only one GlnB-like protein (e.g. cyanobacteria, azotobacter), enabling the tight coordination of carbon and nitrogen metabolisms. Similarly, patterns of sequestration may be a biophysical regulatory mechanism operating widely in other contexts, and understanding its spatiotemporal dynamics in the cell would help link gene and metabolic regulation responses to biophysical feedback to environmental inputs~\cite{alley2008biophysical}.
   
\section*{Author contributions}
AG, MBa, MBu and JS designed the study. AG performed the analysis under the supervision of MBa and MBu. JS designed and coordinated experiments performed by  VB, JGB, FH, and MBe. AG, MBa, JS and MBu wrote the manuscript, with inputs from other authors.    
    
\section*{Acknowledgements}
We thank Mariano Beguerisse-Diaz for his intellectual contributions about parameter fitting. This work was funded by a BBSRC LoLa grant (BB/G020434/1). AG acknowledges funding through a PhD studentship under the BBSRC DTP at Imperial College London (BB/M011178/1). M. Barahona acknowledges funding from the EPSRC (EP/N014529/1 and EP/I032223/1).

\section*{Data availability}
Data supporting this study is available at figshare with DOI: 10.6084/m9.figshare.4880003.

\section*{Conflict of interest}
The authors declare that they have no conflict of interest.

\bibliographystyle{biophysj}
\bibliography{references}
\newpage

\setcounter{figure}{0}
\setcounter{section}{0}
\setcounter{equation}{0}
\setcounter{table}{0}

\renewcommand\thesection{S\arabic{section}}
\renewcommand\thefigure{S\arabic{figure}}
\renewcommand\theequation{S\arabic{equation}}
\renewcommand\thetable{S\arabic{table}}

\begin{center}
{\LARGE Supporting Material: \\ GlnK facilitates the dynamic regulation of bacterial nitrogen assimilation}
\end{center}

\section{Kinetic mechanism for an enzyme with multiple alternative substrates}

Here we present the detailed mathematical formulation of the kinetics of an enzyme with two alternative substrates as depicted on Fig. 4a and described by the following reactions:
\begin{subequations}
\label{eq:ut}
\begin{align}
\varA{GlnB}+\varA{E_1}  &\xrightleftharpoons[a_1]{d_1}  \varA{E_1GlnB} \xrightarrow[]{k_1} \varA{GlnB-U}+\varA{E_1} \label{eq:ut1}\\
\varA{GlnK}+\varA{E_1}  &\xrightleftharpoons[a_2]{d_2}  \varA{E_1GlnK} \xrightarrow[]{k_2} \varA{GlnK-U}+\varA{E_1} \label{eq:ut2}\\
\varA{GlnB-U}+E_2  &\xrightleftharpoons[a_3]{d_3}   \varA{E_2GlnB-U} \xrightarrow[]{k_3}  \varA{GlnB}+ \varA{E_2} \label{eq:ur1}\\
\varA{GlnK-U}+ \varA{E_2}  &\xrightleftharpoons[a_4]{d_4}   \varA{E_2GlnK-U}\xrightarrow[]{k_4} \varA{GlnK}+ \varA{E_2} \label{eq:ur2}
\end{align}
\end{subequations}
where \eqref{eq:ut1} and \eqref{eq:ut2} represent uridylylation, whereas \eqref{eq:ur1} and \eqref{eq:ur2} represent deuridylylation. Here $k_1$, $k_2$, $k_3$ and $k_4$ are the catalytic rate constants, which are functions of the glutamine concentration. Furthermore, $\varA{E_1}$ and $\varA{E_2}$ denote the concentration of the enzyme UT/UR with UT and UR activities respectively. To derive the rate equations corresponding to reactions \eqref{eq:ut}, we assume that complex formation between the substrates and the enzyme occurs much faster than product formation, which is the standard assumption in Michaelis-Menten kinetics \citep{cornish}. Furthermore, since the enzyme active sites are shared between the two substrates, we may treat either substrate as the competitive inhibitor of the other. Using techniques described in section 6.2.1 of  \cite{cornish} we obtain the following system of four equations:
\begin{subequations} \label{eq:ut3}
\begin{align} 
v_\text{UT,GlnB}&=\frac{V_\text{UT,GlnB}\text{[GlnB]}}{K_\text{m,GlnB}(1+\text{[GlnK]}/K_\text{m,GlnK})+\text{[GlnB]}} \\
v_\text{UT,GlnK}&=\frac{V_\text{UT,GlnK}\text{[GlnK]}}{K_\text{m,GlnK}(1+\text{[GlnB]}/K_\text{m,GlnB})+\text{[GlnK]}} \\
v_\text{UR,GlnB-U}&=\frac{V_\text{UR,GlnB-U}\text{[GlnB-U]}}{K_\text{m,GlnB-U}(1+\text{[GlnK-U]}/K_\text{m,GlnK-U})+\text{[GlnB-U]}} \\
v_\text{UR,GlnK-U}&=\frac{V_\text{UR,GlnK-U}\text{[GlnK-U]}}{K_\text{m,GlnK-U}(1+\text{[GlnB-U]}/K_\text{m,GlnB-U})+\text{[GlnK-U]}}
\end{align}
\end{subequations}
where $K_{m,*}=(d_i+k_i)/a_i$, $i \in \{1,2,3,4\}$ are the usual Michaelis constants, and $V_\text{UT,*}=k_p[E_1]$, $p \in \{1,2\}$ and $V_{UR,*}=k_q[E_2]$, $q \in \{3,4\}$ are the maximum enzyme velocities. 

\begin{figure}
    \centering
    \includegraphics[width=0.9\textwidth]{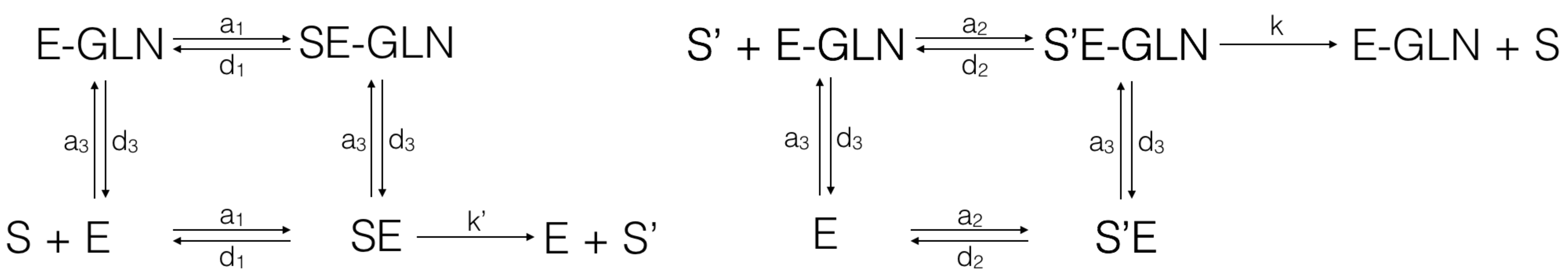}
    \caption{Mechanism for allosteric regulation of UT/UR by glutamine (GLN) based on Fig. 4a. On the left hand side glutamine inhibits the conversion of the substrate $S$ (GlnB, GlnK) to $S'$ (GlnB-U, GlnK-U), whereas on the right hand side GLN activates the conversion of $S'$ to $S$. It is assumed that the substrate binding is unaffected by the presence of GLN.}
    \label{UTmodel}
\end{figure}

In addition to the competing substrates, glutamine (GLN) may also bind to the allosteric side of UT/UR, independently of the substrate binding at a rate $K_\text{GLN}=d_5/a_5$, where $a_5$ and $d_5$ are the association and dissociation constants of GLN. When bound, GLN activates the UR activity and inhibits the UT activity \cite{ninfa1}. We may describe this using the noncompetitive activation and inhibition model (see Section 6.2.3 and Table 6.1 of \cite{cornish}), depicted 
in Fig.~\ref{UTmodel}), which result in an additional term modifying the $V_\text{m,*}$'s of eqs. \eqref{eq:ut3}:
\begin{subequations} \label{eq:ut4}
\begin{align} 
v_\text{UT,GlnB}&=\frac{V_\text{UT,GlnB}\text{[GlnB]}/K_\text{m,GlnB}}{(1+\text{[GlnK]}/K_\text{m,GlnK}+\text{[GlnB]}/K_\text{m,GlnB})}\left( \frac{1}{1+\text{[GLN]}/K_\text{GLN}}\right)\\
v_\text{UT,GlnK}&=\frac{V_\text{UT,GlnK}\text{[GlnK]}/K_\text{m,GlnK}}{(1+\text{[GlnB]}/K_\text{m,GlnB}+\text{[GlnK]}/K_\text{m,GlnK})}\left(\frac{1}{1+\text{[GLN]}/K_\text{GLN}}\right)\\
v_\text{UR,GlnB-U}&=\frac{V_\text{UR,GlnB-U}\text{[GlnB-U]}/K_\text{m,GlnB-U}}{(1+\text{[GlnK-U]}/K_\text{m,GlnK-U}+\text{[GlnB-U]}/K_\text{m,GlnB-U})}\left(\frac{1}{1+K_\text{GLN}/\text{[GLN]}}\right) \\
v_\text{UR,GlnK-U}&=\frac{V_\text{UR,GlnK-U}\text{[GlnK-U]}/K_\text{m,GlnK-U}}{(1+\text{[GlnB-U]}/K_\text{m,GlnB-U}+\text{[GlnK-U]}/K_\text{m,GlnK-U})}\left(\frac{1}{1+K_\text{GLN}/\text{[GLN]}}\right).
\end{align}
\end{subequations}

\section{Kinetic mechanism for an enzyme with competing allosteric effectors}\label{sec:ATase}

The mathematical description of simple inhibition and activation can be found in standard textbooks on reaction \citep{cornish}. However, enzymes with multiple competing allosteric effectors such as AT/AR are rare, therefore we need to give a derivation of such a system. Fig. 4b illustrates graphically an enzyme which can bind two allosteric effectors (inhibitors or activators) and a substrate molecule simultaneously, but their presence do not influence each others' binding. As in Michaelis-Menten kinetics, we assume that the reactions involving complex formation between the enzyme, the substrates and the effectors are in equilibrium independently and at a much shorter timescale than that at which the reaction takes place. 

\begin{figure}
    \centering
    \includegraphics[width=0.9\textwidth]{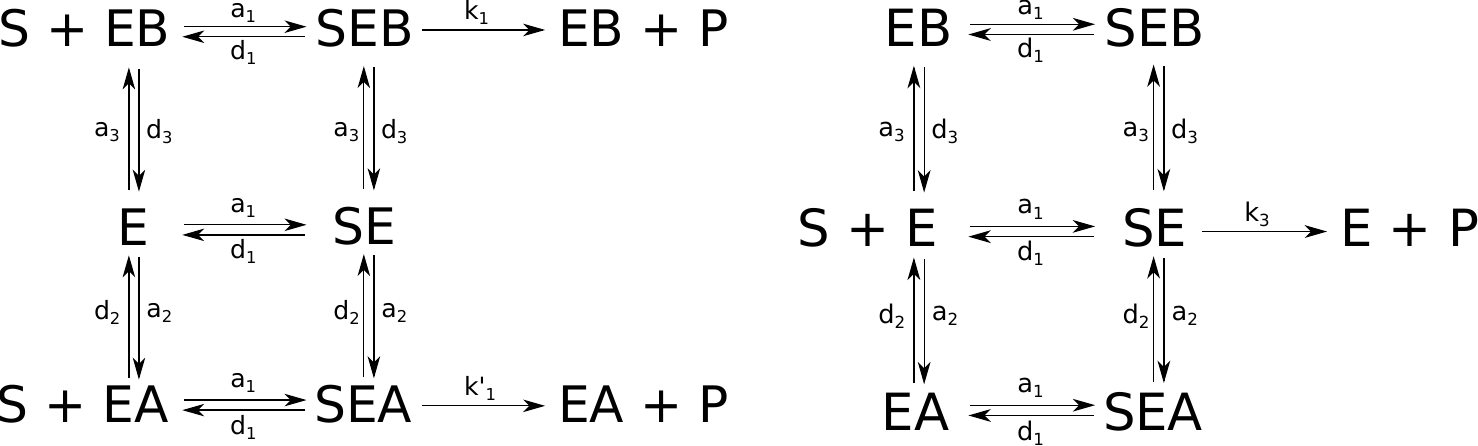}
    \caption{Mechanism for allosteric regulation of AT/AR by competitive effectors based on Fig. 4b. On the left hand side proteins $A$ and $B$ compete for activation of enzyme $E$, whereas on the right hand side they compete for inhibition of enzyme $E$. It is assumed that the substrate binding is unaffected by the presence of the effector.}
    \label{ATmodel}
\end{figure} 

We begin with the case of multiple competing allosteric activators, which can be depicted by the reaction diagram on the left of Fig.~\ref{ATmodel}. Let $E$ denote a general enzyme catalysing the conversion of a substrate $S$ to a product $P$. Assume that $E$ is in a neutral state unless one of the activators $A$ or $B$ (having different affinities) bind to the common allosteric site. When either effector is bound, product formation may take place, however the reaction rate depends on whether $A$ or $B$ is bound. To express the reaction rate mathematically as a function of effector concentrations note that we can define five different complexes whose concentrations are denoted as: $c_1=[SE]$, $ c_2=[SEA]$, $c_3=[EA]$, $c_4=[EB]$, $c_5=[SEB]$. Also, let $s=[S]$, $a=[A]$, $b=[B]$. Then by the law of mass action it follows that:
\begin{subequations}
\label{eq:effectors}
\begin{align} 
(e_0-c_1-c_2-c_3-c_4-c_5)s &=  K_1c_1 \label{subeq1}\\ 
(e_0-c_1-c_2-c_3-c_4-c_5)a &=  K_2c_3 \label{subeq2}\\
(e_0-c_1-c_2-c_3-c_4-c_5)b &=  K_3c_4 \label{subeq3}\\
c_3s &= K_1c_2 \label{subeq4}\\
c_4s &= K_1c_5 \label{subeq5}\\
c_1a &= K_2c_2 \label{subeq6}\\
c_1b &= K_3c_5 \label{subeq7}
\end{align}
\end{subequations}
where $e_0=e+c_1+c_2+c_3+c_4+c_5$ is the total concentration of the enzyme and $K_i=d_i/ a_i$. This is a linear system of equations with eight variables. There are seven equations, but two are linear combinations of the other five (system has rank 5) so we can express the concentration of the complexes as functions of three, which we choose to be $s$, $a$ and $b$. Substituting  \eqref{subeq4}, \eqref{subeq5}, \eqref{subeq6} and \eqref{subeq7} into \eqref{subeq1} we obtain:
\begin{align*} 
\left(e_0-\frac{K_2}{ a}c_2-c_2-\frac{K_1}{ s}c_2-\frac{K_1}{ s}\frac{b}{ K_3}\frac{K_2 }{ a}c_2-\frac{b }{ K_3}\frac{K_2 }{ a}c_2 \right)s &=  K_2\frac{K_1}{ a}c_2 \\ 
\left(e_0-\frac{K_3}{ b}c_5-\frac{a}{ K_2}\frac{K_3}{ b}c_5-\frac{K_1}{ s}\frac{a }{ K_2}\frac{ K_3 }{ b}c_5-\frac{K_1 }{ s}s_5-c_5 \right)a &=  K_3\frac{K_1 }{ b}c_5.
\end{align*}
Rearranging, it follows that:
\begin{align*} 
c_2&=\frac{e_0}{\left(1+\frac{K_1 }{ s}\right)\left(1+\frac{K_2}{ a}+\frac{b}{ K_3}\frac{K_2 }{ a}\right)} \\
c_5&=\frac{e_0}{\left(1+\frac{K_1 }{ s}\right)\left(1+\frac{K_3}{ b}+\frac{a}{ K_2}\frac{K_3 }{ b}\right)}.
\end{align*}
Hence the catalytic rate of the enzyme is:
\begin{align} 
v=k_1c_2+k'_1c_5=\frac{1}{\left(1+\frac{K_1 }{ s}\right)} \left[ \frac{k_1e_0}{(1+\frac{K_2}{ a}+\frac{b}{ K_3}\frac{K_2 }{ a})}+\frac{k'_1e_0}{(1+\frac{K_3}{ b}+\frac{a}{ K_2}\frac{K_3 }{ b})} \right].
\end{align}
Hence comparing with the case of a single activator (see Cornish-Bowden 2013, pg. 152-157) we obtain that there the effect of the two activators are almost additive except for an additional mixed term in the denominator.

We proceed similarly to obtain a relationship for the case of multiple competing allosteric inhibitors (see right of Fig. \ref{ATmodel}). Using rapid equilibrium assumption we obtain a set of equations identical to system (\ref{eq:effectors}). Here we need to solve for complex $c_1$, so we write:
\begin{align*} 
\left(e_0-c_1-\frac{b }{ K_3}c_1- \frac{K_1}{ s}\frac{b}{K_3}c_1-  \frac{K_1 }{ s}\frac{a}{ K_2}c_1- \frac{a}{ K_2}c_1\right)s&=K_1c_1, 
\end{align*}
from which it follows that the enzyme velocity is:
\begin{align} 
v=k_1c_1= \frac{e_0}{\left(\frac{K_1}{s}+1 \right) \left(1+\frac{a }{K_2}+\frac{b }{K_3} \right)}.
\end{align}
So similarly to the case of single inhibitor we obtain that there is a reduction of the maximum velocity of the reaction leaving the $K_1$ unchanged.

\clearpage
\section{Global sensitivity analysis of the model}
\label{sec:sensitivity}

To assess the significance of the model parameters we conducted a sensitivity analysis. The objective of the latter is to quantify the change in the system output to a small perturbation in one or a combination of parameters, while the others are kept constant. Since the outputs are time series, the distance between two time series if of interest, therefore conventional (local) sensitivity analysis techniques, which are only concerned with variations in the steady state are not adequate. Instead, here we use the eFAST method \citep{saltelli} to assess parameter sensitivities, which is a global sensitivity analysis technique, and is available in the Systems Biology Toolbox 2 for Matlab \citep{schmidt}. In essence, global sensitivity analyses are concerned with sampling the parameter space - usually randomly, according to some optimal strategy - in the vicinity of the fitted parameters and estimating the resulting variation in the output. The latter is usually measured by the pointwise difference between the nominal and perturbed output time series integrated over the time range of interest. The eFAST method in particular samples the parameter space along such a trajectory that allows the model output to be expressed as a Fourier series. Then the variance in the output can be decomposed into a sum of terms involving Fourier coefficients, which can be estimated using Monte Carlo techniques. In our analysis we used a 10\% perturbation from the nominal parameters and $10^5$ samples.

\begin{figure}
	\centering
	\includegraphics[width=0.5\textwidth]{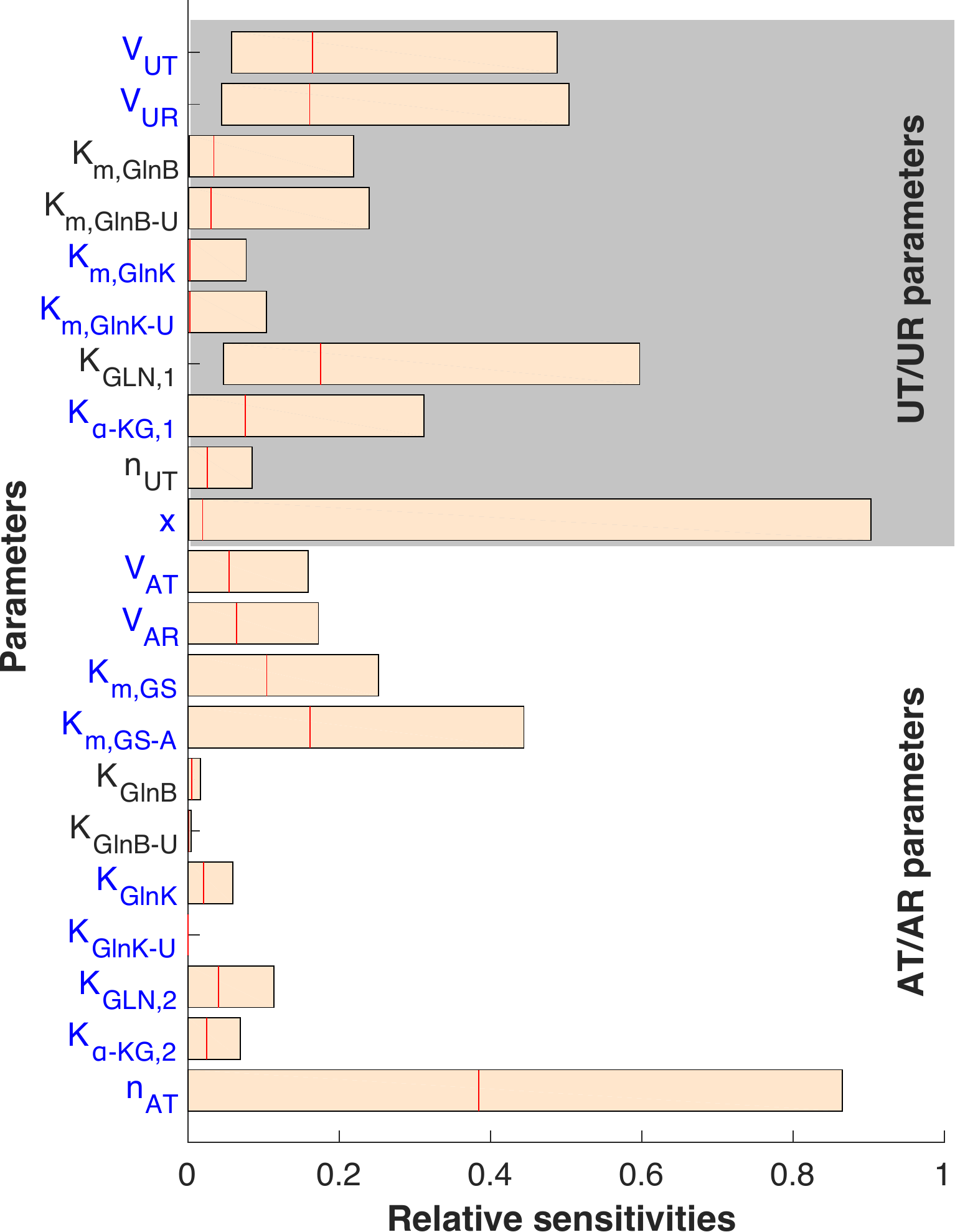}
	\caption{Relative sensitivity of variables to model parameters obtained using eFAST global sensitivity analysis~\citep{saltelli,schmidt}. The fitted parameters are marked in blue. Bars show the maximum and minimum sensitivity of the individual variables (GlnB-U, GlnK-U and GS-A) obtained for $10^5$ samples from the parameter space with a maximum 10\% deviation from the fitted parameters. Red line indicates the median value.}
	\label{par_sens}
\end{figure}

Fig.~\ref{par_sens} shows the relative sensitivities of the model parameters. Since GlnB/GlnB-U and GlnK/GlnK-U are important effectors of the AT/AR enzyme, GS-A levels will depend on the parameters that are specific to uridylylation. However, GS-A levels do not feed back into the model since we used glutamine concentration as a driver. Hence the parameters specific to adenylylation do not affect GlnB-U and GlnK-U states. Hence we grouped the parameters according to whether they are involved in uridylylation or adenylylation. As expected, the most sensitive parameters are the maximum enzyme velocities, Michaelis constants and the Hill-coefficient of the AT/AR reactions ($n_\text{AT}$), which has a large effect on setting the basal GS-A levels. Another parameter of high sensitivity is the percentage sequestration, $x$, which confirms the importance of modelling AmtB-GlnK complex formation. The parameters of lowest relative sensitivity are the activation constants $K_\text{GlnB}$, $K_\text{GlnB-U}$, hence we fixed these from the literature (see Table I in the paper). Although $K_\text{GlnK-U}$ was also found to be relatively insensitive, which is commensurate with the wide sample distribution in Fig.~\ref{pardistr}, we could not find parameters in the literature measured under similar conditions to our experiments. We leave these at their fitted values, because they are highly influential for our predictions.

\clearpage
\section{Evolutionary Monte Carlo optimisation algorithm for parameter fitting}

The parameters were fitted using the 'Squeeze and Breathe' evolutionary optimisation method \citep{beguerisse}. Let $\mathbf{X}(t)=\{x_1(t),\dots,x_d(t)\}$ denote the state of the system with $d$ variables at time $t$, which in our case are the concentrations of GlnB, GlnK and GS. The evolution of these variables is described by a set of ODEs, $\dot{\mathbf{X}}=f(\mathbf{X},t;\mathbf{\theta})$, where $\theta=\{\theta_1,\dots \theta_n\}$ is a set of $n$ parameters. In essence, the objective of the parameter optimisation is to find parameters $\theta$ such that the distance between the solution $\mathbf{X}$ and an experimental dataset $\mathcal{D}=\{\widetilde{\mathbf{X}}(t_i)|i=1,\dots,m\}$ of $m$ observations is minimised. To account for measurement error we define the following cost function, weighted by the standard error of the measurements: 
\begin{align*}
	E_\mathcal{D}(\theta)=\min_{\theta}\left\{ \sum_{j=1}^m{\| ( \mathbf{X}(t_i;\mathbf{\theta})-\widetilde{\mathbf{X}}(t_i) )/ SE_i\|} \right\},
\end{align*}
where $\| .\|$ is the Euclidean norm. Since the cost function depends on all $n$ parameters, its value lies in an $n$ dimensional space. The optimal parameters will be the coordinates corresponding the global minimum of $E_\mathcal{D}(\theta)$ over all possible parameters. The cost function $E_\mathcal{D}(\theta)$ defines a very rough landscape and as a result optimisation methods may get stuck in a local minimum. In fact, an algorithm which guarantees to find the best parameters does not exist. Therefore the objective of the parameter optimisation is to explore a large portion of this space to get as close to the global minimum as possible. The Squeeze and Breathe algorithm achieves this by first running local optimisation around random samples in the parameter space. These are then ranked according to optimality and culled keeping only the best few. The culled set is then used to obtain a posterior sampling distribution. The process is repeated until the difference between subsequent posterior distributions is small. 

\begin{figure}
	\centering
    \includegraphics[width=0.7\textwidth]{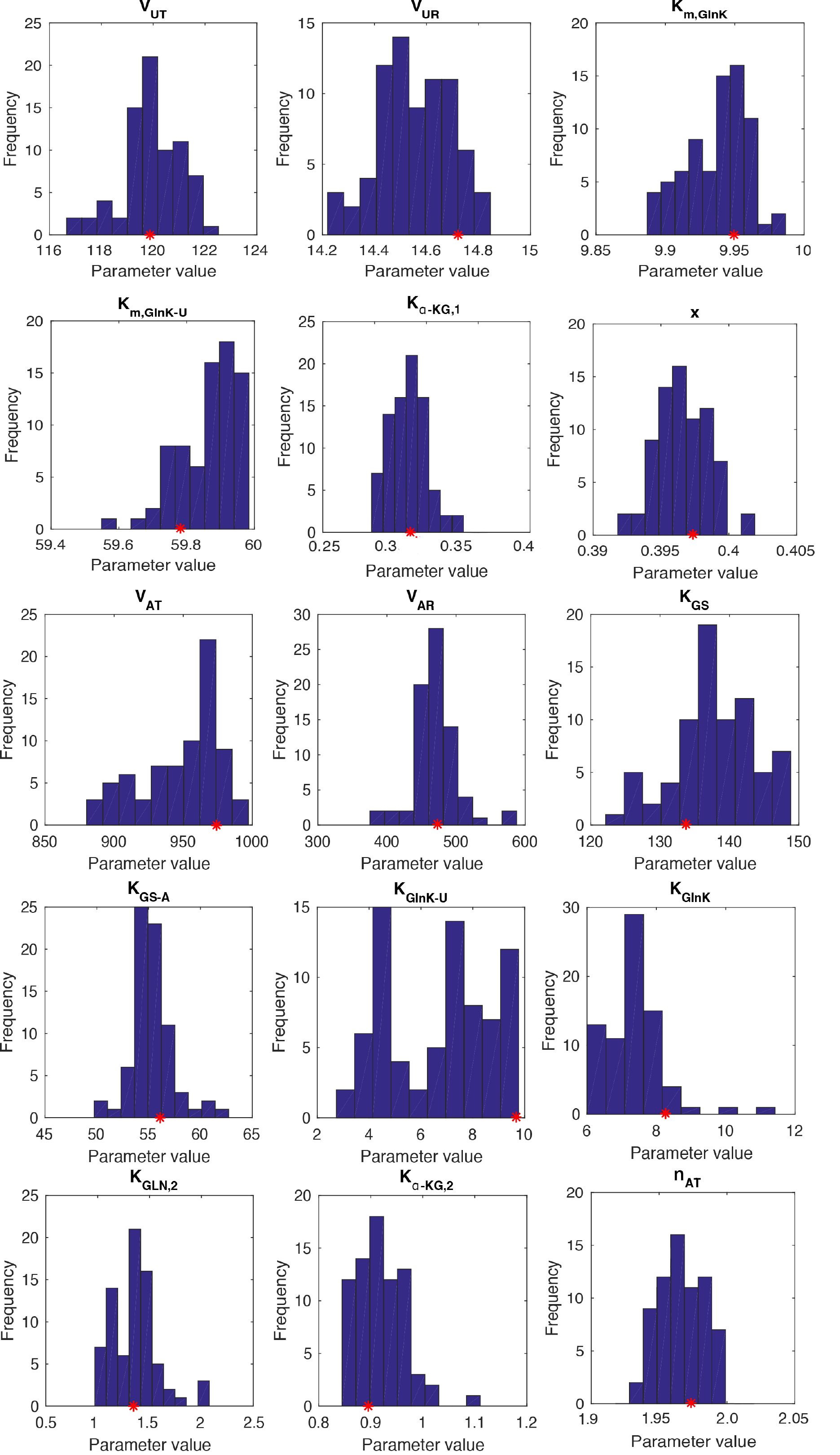}
    \caption{Histograms of fitted parameters obtained through the Squeeze-and-Breathe algorithm~\citep{beguerisse}. Red asterisks indicate the parameter combination with the lowest cost function.}
    \label{pardistr}
\end{figure}

The fitted parameters are shown in Table I in the main paper and the histograms from the parameter sampling are shown in Fig. \ref{pardistr}. The histograms show the number of times the parameter fitting algorithm converged to the particular parameter values, whereas the red asterisk show the parameter value with the minimum cost function. A narrow distribution around a fitted parameter reflects a well defined minimum in the explored parameter range, indicating that the fitted model has higher sensitivity to the variations of these parameters. On the other hand a wide distribution shows that the cost function landscape is shallow or contains many local minima. This indicates that the model might be less sensitive to the corresponding parameters. Fig. \ref{pardistr} show that most parameter have a narrow spread. An exception is $K_\text{GlnK-U}$, which show high variation, indicating low sensitivity of the model to this parameter (confirmed by sensitivity analysis). In the main paper we argue, that the low sensitivity of $K_\text{GlnK-U}$ in the WT model has biological origins, since the GlnK protein is less potent compared to GlnB. However, we do not $K_\text{GlnK-U}$ to a value found in the literature, since this parameter will be essential for accurate prediction of the GS adenylylation levels in the mutant strain containing no GlnB. The parameters to be fitted were chosen also based on the information provided by a global sensitivity analysis of the model presented in Section~\ref{sec:sensitivity}. 

\clearpage

\section{Model selection: uridylylation reactions with and without sequestration}

We compared the uridylylation models with and without GlnK sequestration (Eq. 4 in the main text) using different model selection criteria. All criteria unanimously selected the model with sequestration. 

Our model is fitted to the WT time-series (training dataset). The fitted model is then used to predict the dynamical responses for the two mutants (test datasets) without using the measurements. The model with sequestration provides an improved fit both for the training dataset and an improved prediction of the test datasets, yet it contains 6 parameters (instead of 4 parameters for the model without sequestration). 

To make the comparison precise, we obtain the statistical significance of the models with and without sequestration based on two information criteria:
\begin{itemize}
\item Akaike Information Criterion corrected for small datasets: $$\text{AICc}=n\log(\text{RSS}) + \frac{2kn}{n-k-1}$$ 
\item Bayesian Information Criterion: $$\text{BIC} = n\log(\text{RSS}) + k\log(n)$$
\end{itemize}
Here $n$ is the number of data points (24 using the WT and the $\Delta$\textit{glnB} datasets); $k$ is the number of fitted parameters (4 without sequestration and 6 with sequestration); and RSS is the residual sum-of-squares deviation of the model from the data (Eq. 7 ). 

Both criteria compare models based on their goodness-of-fit and penalise the number of parameters. The AICc criterion is theoretically more appropriate (since it compares the model to the true model), but has a bias toward models with higher complexity. As a further confirmation, we also computed the BIC criterion, which puts a higher penalty on the number of model parameters, and would thus favour the lower complexity model. 

Table~\ref{modelselect} shows that the model with sequestration is selected according to both the AICc and BIC criteria, as shown by lower values of both criteria for the model with sequestration. 

\begin{table}[h!]
\centering
\caption{Information criteria for model selection of uridylylation model with and without sequestration using the WT and $\Delta$\textit{glnB} data ($n=24$ points)}
\vspace*{5mm}
\label{modelselect}
\begin{tabular}{|l | c | c |c |c |}
\hline 
\textbf{Model} & \textbf{Number of fitted parameters ($k$)} & \textbf{RSS} & \textbf{AICc} & \textbf{BIC} \\ 
\hline
  without sequestration & 4 & 60.3 & 108.5 & 111.1  \\ \hline
  with sequestration      & 6 & 21.7 & \textbf{90.8} & \textbf{92.9}  \\
   \hline 
\end{tabular}
\end{table}

\clearpage
\section{Comparison to the Straube model of GlnB uridylylation}

\begin{figure}
    \centering
    \includegraphics[width=0.55\textwidth]{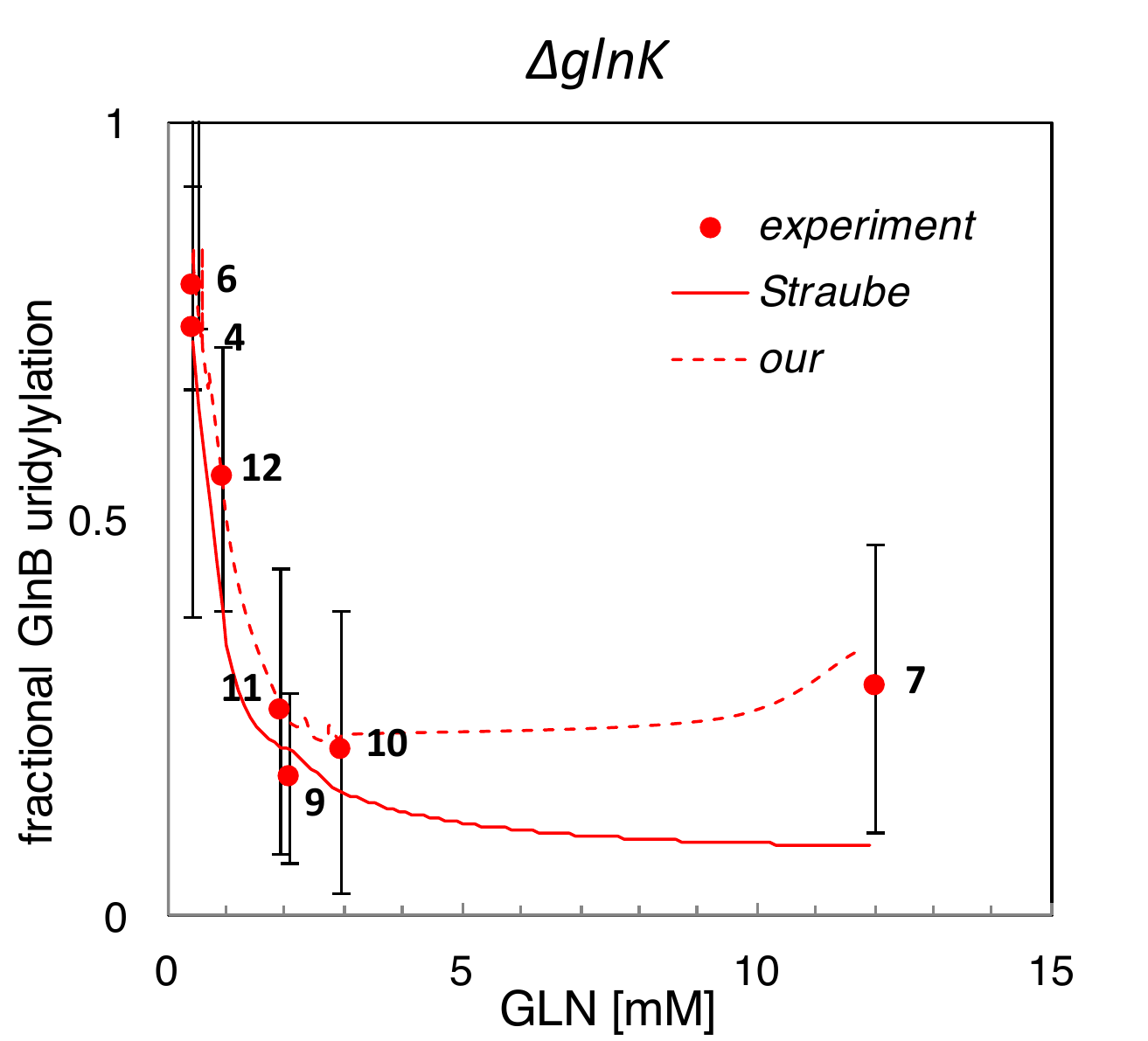}
    \caption{Measured and fitted dynamics of fractional GlnB uridylylation against glutamine for our model and the Straube model, normalised to total concentrations (Fig. 3 in paper). Error bars represent mean squared error of two independent measurements (total and modified protein) with three replicates each. For both figures and models we used parameters from the in vitro reconstituted GlnB-UT/UR system. Numbers correspond to sample points on other figures. Both models can fit the data equally well for all glutamine conditions in the $\Delta$\textit{glnK} strain except 30s after ammonium upshift (point 7) where the steady state Straube model underestimates GlnB uridylylation.}
    \label{straube}
\end{figure}

The enzyme UT/UR has two active sites \citep{Zhang}, catalysing uridylylation (UT) of GlnB/GlnK and deuridylylation (UR) of GlnB-U and GlnK-U. Having two active sites both with two distinct substrates means there are four different ways in which ternary complexes can form. To test whether these contribute significantly to product formation in the UT and UR reactions we compared the Straube model \citep{straube}, which describes GlnB uridylylation and includes the ternary complex between UT/UR, GlnB and GlnB-U, and another model, where the contribution of the ternary complex is ignored. The latter is equivalent to the classical Goldbeter-Koshland model \citep{gk} that assumes Michaelis-Menten kinetics with allosteric inhibition and activation by glutamine for the UR and UT reactions respectively. For both models we used literature parameters from the \textit{in vitro} reconstituted GlnB-UT/UR system \citep{ninfa3} except for the $V_{max}$, which had to be fitted. To make the best comparison we used the fact  that $V_\text{max}=ke$, where $k$ is the catalytic rate and $e$ is the concentration of the enzyme, fixed the ratio between $V_{UT}/V_{UR}=k_{UT}/k_{UR}$ using literature values of $k_{UT}$ and $k_{UR}$ \citep{ninfa3} and fitted only one parameter, corresponding to $e$. As Fig.~\ref{straube} shows, both models produce indistinguishable results under most conditions except 30s after upshift where the Straube model, being a steady state model, performs worse due to strong transient dynamics. This suggests that the contribution of the ternary complexes are not likely to be significant. 

Despite performing well on GlnB uridylylation using literature parameters, the Straube model could only produce an adequate fit to the GlnK data in the $\Delta$\textit{glnB} strain when the basal UT activity (which is one of its parameters) was as high as 50\%. We found this after an extensive parameter search using a non-linear least squares fitting procedure. This is much higher than the 1\% basal UT activity reported in a previous in vitro study with GlnB and UT/UR \citep{ninfa3}. Furthermore, due to the number of complexes this model cannot be extended to treat GlnB and GlnK at the same time.

The high basal UT activity for GlnK in the Straube model highlights the asymmetry in the system with respect to glutamine. In other words, glutamine binding to UT/UR activates the UR activity and inhibits the UT activity by different relative amounts.  We could account for this in our model by defining two different $K_d$ for the activation of UR and inhibition of UT reactions by glutamine. 

\clearpage
\section{Additional Supplementary Figures and Tables}

\begin{figure}[h!]
    \centering
    \includegraphics[width=\textwidth]{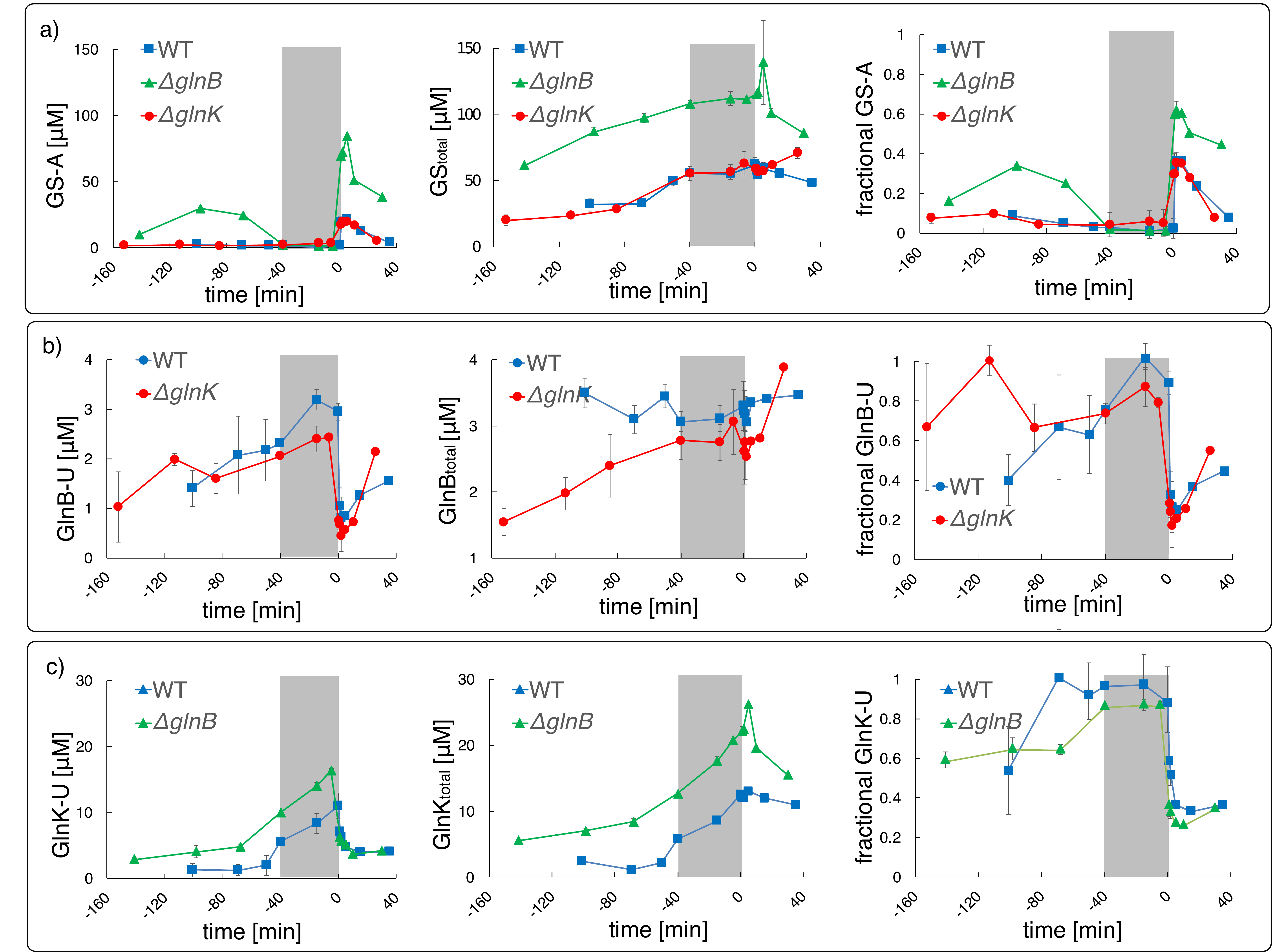}
    \caption{PTM state, total concentration and fractional PTM state of a) GS b) GlnB and c) GlnK proteins during run-out, starvation and a subsequent ammonium upshift at $t=0 s$. Starvation is marked by the grey area. The symbols and error bars represent averages $\pm$s.e (n=3). The lines are guides to the eye.}
\end{figure}

\begin{figure}[h!]
    \centering
    \includegraphics[width=0.5\textwidth]{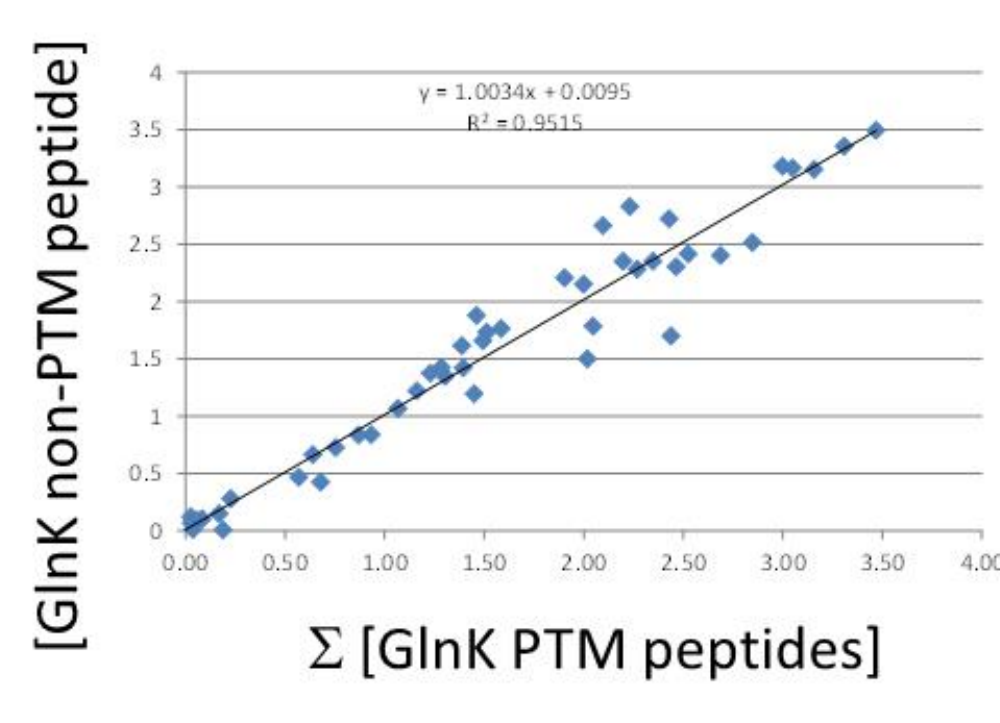}
    \caption{Correlation between GlnK concentration derived from non-uridylylated peptides and that derived independently from the sum of uridylylated and non-uridylylated peptide GAEYSVNFLPK.}
\end{figure}

\begin{table}
\centering
\caption{MRM-MS signals of GlnK unlabelled/labelled signature peptides}
\vspace{5mm}
\label{pept}
\resizebox{\textwidth}{!}{%
\begin{tabular}{ccccccc}
\textbf{Protein} & \textbf{Peptide} & \textbf{\begin{tabular}[c]{@{}c@{}}Internal \\ standard (Y/N)\end{tabular}} & \textbf{Q1} & \textbf{Q3} & \textbf{Retention} & \textbf{Collision energy} \\ \hline
GlnK-1 & GAEYSVNFLPK-1a & N & 612.8 & 804.5 & 35.8 & 30 \\
GlnK-1 & GAEYSVNFLPK-1b & N & 612.8 & 967.5 & 35.8 & 30 \\
GlnK-1 & GAEYSVNFLPK-1c & N & 612.8 & 244.2 & 35.8 & 30 \\
GlnK-1 & GAEYSVNFLPK-1a-is & Y & 616.8 & 812.5 & 35.8 & 30 \\
GlnK-1 & GAEYSVNFLPK-1b-is & Y & 616.8 & 975.5 & 35.8 & 30 \\
GlnK-1 & GAEYSVNFLPK-1c-is & Y & 616.8 & 252.2 & 35.8 & 30 \\ \hline
GlnK(u)-1 & GAEY(U)SVNFLPK-1a & N & 765.9 & 244.2 & 33.2 & 30 \\
GlnK(u)-1 & GAEY(U)SVNFLPK-1b & N & 765.9 & 804.5 & 33.2 & 30 \\
GlnK(u)-1 & GAEY(U)SVNFLPK-1a-is & Y & 769.9 & 252.2 & 33.2 & 30 \\
GlnK(u)-1 & GAEY(U)SVNFLPK-1b-is & Y & 769.9 & 812.5 & 33.2 & 30 \\ \hline
GlnK-2 & IFVAELQR-2a & N & 488.3 & 715.4 & 31.8 & 30 \\
GlnK-2 & IFVAELQR-2b & N & 488.3 & 616.3 & 31.8 & 30 \\
GlnK-2 & IFVAELQR-2c & N & 488.3 & 862.5 & 31.8 & 30 \\
GlnK-2 & IFVAELQR-2a-is & Y & 493.3 & 725.4 & 31.8 & 30 \\
GlnK-2 & IFVAELQR-2b-is & Y & 493.3 & 626.3 & 31.8 & 30 \\
GlnK-2 & IFVAELQR-2c-is & Y & 493.3 & 872.5 & 31.8 & 30 \\ \hline
\end{tabular}%
}
\end{table}

\clearpage

\bibliographystyle{biophysj}
\bibliography{references}

\end{document}